\documentclass[12pt,a4paper]{article}
\pdfoutput=1

\usepackage{geometry}
\usepackage{amsmath}
\usepackage{amsmath}
\usepackage{amssymb}
\usepackage[dvips]{graphicx}
\usepackage{cite}
\usepackage{pstricks}
\usepackage{bm}
\usepackage{pbox}
\usepackage{placeins}
\usepackage{graphicx}
\usepackage{caption}
\usepackage{subcaption}
\usepackage[T1]{fontenc}
\usepackage{footnote}
\usepackage{pdfpages}
\usepackage{hhline}
\usepackage{multirow}
\usepackage{multicol}
\usepackage{enumitem}
\usepackage{multirow}
\usepackage{amsmath}
\usepackage{relsize}
 \usepackage{graphicx}
\usepackage[toc,page]{appendix}
\usepackage[english]{babel}
\allowdisplaybreaks

\usepackage{slashed}
\usepackage{cases}

\usepackage{empheq}

\renewcommand{\thetable}{\Roman{table}}

\definecolor{MyDarkBlue}{rgb}{0.1, 0.3, 0.8} %defining the color 'MyDarkBlue'
\definecolor{SBlue}{rgb}{0.2, 0.4, 0.4} %defining the color 'MyDarkBlue'
\definecolor{MyLightBlue}{rgb}{0.22,0.51,0.99}
\definecolor{MyGreen}{rgb}{0.0, 0.5, 0.3}
\definecolor{BrickRed}{rgb}{0.8, 0.25, 0.33}
\usepackage[colorlinks=true,linkcolor=blue,citecolor=MyDarkBlue,
urlcolor=BrickRed,bookmarksnumbered=true,bookmarksopen]{hyperref}
\hypersetup{colorlinks, citecolor=SBlue,linkcolor=MyGreen, urlcolor=MyDarkBlue}

\begin{document}
\vspace*{-0.2in}
\begin{flushright}
{ OSU-HEP-20-07}
\end{flushright}
\vspace{0.1cm}
\begin{center}
{\Large \bf
%Large Neutrino Transition Magnetic Moment and Tiny Mass Generation in the %Light of XENON1T Electron Recoil Excess
%Neutrino Magnetic Moment in Light of Recent Experiments
%Neutrino Masses and Large  Magnetic Moment  \\  in Light of Recent Experiments
Large Neutrino Magnetic Moments  \vspace{0.08in}\\   in the Light of Recent Experiments
%Large Neutrino Magnetic Moment and Neutrino Mass  \vspace{0.08in}\\  Generation in Light of Recent Experiments
}
\end{center}
\renewcommand{\thefootnote}{\fnsymbol{footnote}}
\begin{center}
{
{}~\textbf{K.S. Babu$^1$}\footnote{ E-mail: \textcolor{MyDarkBlue}{babu@okstate.edu}},
{}~\textbf{Sudip Jana$^2$}\footnote{ E-mail: \textcolor{MyDarkBlue}{sudip.jana@mpi-hd.mpg.de}}
{}~\textbf{and Manfred Lindner$^2$}\footnote{ E-mail: \textcolor{MyDarkBlue}{lindner@mpi-hd.mpg.de}}
}
\vspace{0.5cm}
{
\\\em $^1$Department of Physics, Oklahoma State University, Stillwater, OK 74078, USA
\\
$^2$Max-Planck-Institut f{\"u}r Kernphysik, Saupfercheckweg 1, 69117 Heidelberg, Germany
} 
\end{center}

%\vspace{0.6cm}
\renewcommand{\thefootnote}{\arabic{footnote}}
\setcounter{footnote}{0}
\thispagestyle{empty}

%%%%%%%%%%%%%%%%%%%%%%%%%%%%%%%%%%%%%%%%%%%%%%%
%%%%%%%%%%%%%%%%%%%%%%%%%%%%%%%%%%%%%%%%%%%%%%%
\begin{abstract}

The excess in electron recoil events reported recently by the XENON1T experiment may be interpreted as evidence for a sizable transition magnetic moment $\mu_{\nu_e\nu_\mu}$ of Majorana neutrinos.  We show the consistency of this scenario when a single component transition magnetic moment takes values $\mu_{\nu_e\nu_\mu} \in(1.65 - 3.42) \times 10^{-11} \mu_B$. 
Such a large value typically leads to unacceptably large neutrino masses. In this paper we show that new leptonic symmetries can solve this problem and demonstrate this with several examples. We first revive and then propose a simplified model based on $SU(2)_H$ horizontal symmetry. Owing to the difference in their Lorentz structures, in the $SU(2)_H$ symmetric limit, $m_\nu$ vanishes while $\mu_{\nu_e\nu_\mu}$ is nonzero.  Our simplified model is based on an approximate $SU(2)_H$, which we also generalize to a three family $SU(3)_H$-symmetry.  Collider and low energy tests of these models are analyzed.  We have also analyzed implications of the XENON1T data for the Zee model and its extensions which naturally generate a large  $\mu_{\nu_e\nu_\mu}$ with suppressed $m_\nu$ via a spin symmetry mechanism, but found that the induced $\mu_{\nu_e\nu_\mu}$ is not large enough to explain recent data.  Finally, we suggest a mechanism to evade stringent astrophysical limits on neutrino magnetic moments arising from stellar evolution by inducing a medium-dependent mass for the neutrino.

\end{abstract}

\newpage
\setcounter{footnote}{0}

{
  \hypersetup{linkcolor=black}
  \tableofcontents
}
\newpage

%%%%%%%%%%%%%%%%%%%%%%%%%%%%%%%%%%%%%%%%%%%%%%%
%%%%%%%%%%%%%%%%%%%%%%%%%%%%%%%%%%%%%%%%%%%%%%%
\section{Introduction}\label{SEC-01}
%%%%%%%%%%%%%%%%%%%%%%%%%%%%%%%%%%%%%%%%%%%%%
%%%%%%%%%%%%%%%%%%%%%%%%%%%%%%%%%%%%%%%%%%%%%
The XENON collaboration has recently performed a search for new physics with low-energy electronic recoil data recorded with the XENON1T detector and reported an excess of events over the known backgrounds in the recoil energy range $(1-7)$ keV, peaked around 2.5 keV \cite{Aprile:2020tmw}. This excess, observed with an unprecedented low background rate of (76 $\pm$ 2) events/(tonne $\times$ year $\times$ keV) between $(1-30)$ keV, and an exposure of 1042 kg $\times$ 226.9 days, is quite intriguing. One possible explanation of these anomalous events would be the presence of a sizable neutrino magnetic moment. Within this interpretation, the signal is favored over background at $3.2 \sigma$ significance.  The preferred range of an effective neutrino magnetic moment is  $\mu_{\nu} \in(1.4,2.9) \times 10^{-11} \mu_{B}$ at 90 \% \text { C.L. } \cite{Aprile:2020tmw}.  This excess is also consistent with a solar axion signal, and more conservatively, with a Tritium background in the detector that is unaccounted for \cite{Aprile:2020tmw}. 

Here we wish to explore the large magnetic moment interpretation of the observed XENON1T excess. We interpret the anomaly in terms of a transition magnetic moment $\mu_{\nu_e \nu_\alpha}$ (in flavor basis, where $\alpha = \mu$ or $\tau$) of Majorana neutrinos.  (Majorana neutrinos, being self-conjugate fields, cannot have intrinsic magnetic moments, but can possess transition moments.) Our analysis shows consistency with a single component  transition magnetic moment in the range $\mu_{\nu_e\nu_\mu} \in (1.65 -  3.42) \times 10^{-11} \mu_B$.

In this paper we show that new  symmetries acting on the lepton sector can render 
%Dirac or Majorana
neutrino magnetic moment of order $10^{-11}\mu_B$ compatible with the known neutrino masses. In the absence of additional symmetries (and without severe fine-tuning) one would expect neutrino masses several orders of magnitude larger than their measured values. The main reason for this expectation is that the magnetic moment and the mass operators are both chirality flipping, which implies that by removing the photon line from the loop diagram that induces $\mu_\nu$ one would generate a neutrino mass term.
This would lead to the naive estimate of $m_\nu$ originating from such diagrams given by
\begin{equation}
    m_\nu \sim \frac{ \mu_\nu}{\mu_B} \,\frac{M^2}{2 m_e},
    \label{order}
\end{equation}
where $M$ represents the mass of a heavy particle circling inside the loop diagram.  Since the photon is emitted from an internal line to induce a magnetic moment operator, at least some of the particles inside the loop must be electrically charged.  Experimental limits show that any such charged particle should be heavier than about 100 GeV, in which case Eq. (\ref{order}) would lead to $m_\nu \sim 0.1$ MeV, some six orders of magnitude larger than the observed masses.\footnote{There is an exception for $M$ being large: If the internal particles are milli-charged, direct experimental limits won't exclude them from being light. Even in this case, owing to other experimental constraints on milli-charged particles, the maximum $\mu_\nu$ that can be induced is $\mu_\nu \sim 10^{-15}\mu_B$ \cite{Lindner:2017uvt}.}

This magnetic moment--mass conundrum was well recognized three decades ago when there was great interest in explaining the apparent time variation of solar neutrino flux detected by the Chlorine experiment in anti-correlation with the Sun-spot activity \cite{Davis:1988gd,Davis:1990fb}.  Such a time variation could be explained if the neutrino has a magnetic moment of order $(10^{-11}-10^{-10}) \mu_B$ which would lead to spin-flip transition inside the solar magnetic field \cite{Cisneros:1970nq,Okun:1986na}. Such transitions could even undergo a matter enhanced resonance \cite{Lim:1987tk,Akhmedov:1988uk}.  While this explanation of the solar neutrino data has faded with the advent of other experiments, in the late 1980's and early 1990's there were significant theoretical activities that addressed the compatibility of a large neutrino magnetic moment with a small mass. These discussions become relevant today, if the XENON1T anomaly is indeed a signal of neutrino magnetic moment. 

In this paper we revive and extend mechanisms for enhancing neutrino magnetic moments based on a horizontal $SU(2)_H$ symmetry \cite{Voloshin:1987qy,Babu:1989wn}.  In the limit of this symmetry, owing to differences in the Lorentz structures in the operators, the neutrino mass vanishes while the magnetic moment does not \cite{Babu:1990wv,Ecker:1989ph,Babu:1989px,Chang:1990uga,Leurer:1989hx,Choudhury:1989pw,Babu:1992vq,Babu:1991zh}.  We propose a simplified model based on approximate $SU(2)_H$ symmetry that induces sufficiently large neutrino magnetic moment to explain the XENON1T excess. While the old models almost always relied on exact symmetries, here we show that an {\it approximate} $SU(2)_H$ is sufficient, with explicit breaking of the symmetry provided by the electron and muon masses.  Thus,  new models can be realized with fewer particles, making them simpler.  We also propose an extension of the $SU(2)_H$ symmetry to a three-family $SU(3)_H$ which has the desired property of suppressing neutrino mass while generating large magnetic moment.  Collider and low energy constraints of these models will be analyzed and future tests outlined. A distinct signature of these models for the LHC is the presence of neutral scalars decaying into $\ell^\pm \tau^\mp$ with masses not exceeding a TeV.
We also revisit models with a spin symmetry argument \cite{Barr:1990um,Babu:1992vq} that enhances magnetic moment, but find that these models do not generate large enough $\mu_\nu$ in order to explain the XENON1T data.

Large neutrino magnetic moments are strongly constrained by stellar evolution, since the photon, which has a plasma mass in these surroundings can decay to neutrinos.  The most stringent limit arises from the energy loss of red giant branch in globluar clusters, which require $\mu_\nu < 4.5 \times 10^{-12} \mu_B$ \cite{Viaux:2013lha}.  This value is a factor of 5 below what is needed to explain the XENON1T anomaly. We show here that by invoking interactions of the neutrinos with a light scalar, such plasmon decays may be kinematically suppressed, as the neutrino acquires a medium-dependent mass which is larger than the plasmon mass.

This paper is organized as follows.
In Sec.~\ref{SEC-02} we provide our fit to the XENON1T data in terms of a single component transition magnetic moment $\mu_{\nu_e\nu_\mu}$ and we give a short overview other experimental information.
In Sec.~\ref{SEC-03} we discuss general theoretical aspects of neutrino magnetic moments and point out how new symmetries can explain a large value. Here we review the $SU(2)_H$ symmetric mechanism and the spin symmetry argument to suppress neutrino mass relative to its magnetic moment.
In Sec.~\ref{SEC-04} we present a concrete and simplified model based on an approximate $SU(2)_H$. We carry out a detailed phenomenological analysis of the model in Sec.~\ref{SEC-05}.
In Sec. \ref{SEC-06} we extend the symmetry to an approximate $SU(3)_H$. 
In Sec.~\ref{SEC-07} we analyze magnetic moments in the Zee model and its extensions, where we show their inadequacy to explain XENON1T data. 
In Sec.~\ref{SEC-08}, we suggest a mechanism to evade the astrophysical limits on neutrino magnetic moments.
Finally, we conclude in Sec. \ref{SEC-09}.

%%%%%%%%%%%%%%%%%%%%%%%%%%%%%%%%%%%%%%%%%%%%%%%
%%%%%%%%%%%%%%%%%%%%%%%%%%%%%%%%%%%%%%%%%%%%%%%
\section{Neutrino  magnetic moments: the experimental situation}\label{SEC-02}
%%%%%%%%%%%%%%%%%%%%%%%%%%%%%%%%%%%%%%%%%%%%%
%%%%%%%%%%%%%%%%%%%%%%%%%%%%%%%%%%%%%%%%%%%%%
In this section we briefly summarize the current status of neutrino magnetic moment searches.  First we show the consistency of interpreting the XENON1T excess in terms of a single component transition magnetic moment of $\nu_e$. Then we summarize the experimental status on neutrino magnetic moments from reactor and accelerator neutrinos as well as from astrophysics. 

%%%%%%%%%%%%%%%%%%%%%%%%%%%%%%%%%%%%%%%%%%%%%%%
%%%%%%%%%%%%%%%%%%%%%%%%%%%%%%%%%%%%%%%%%%%%%%%
\subsection{XENON1T }
%%%%%%%%%%%%%%%%%%%%%%%%%%%%%%%%%%%%%%%%%%%%%
%%%%%%%%%%%%%%%%%%%%%%%%%%%%%%%%%%%%%%%%%%%%%

\begin{figure}[h!]
    \centering
    \includegraphics[width=.45\textwidth]{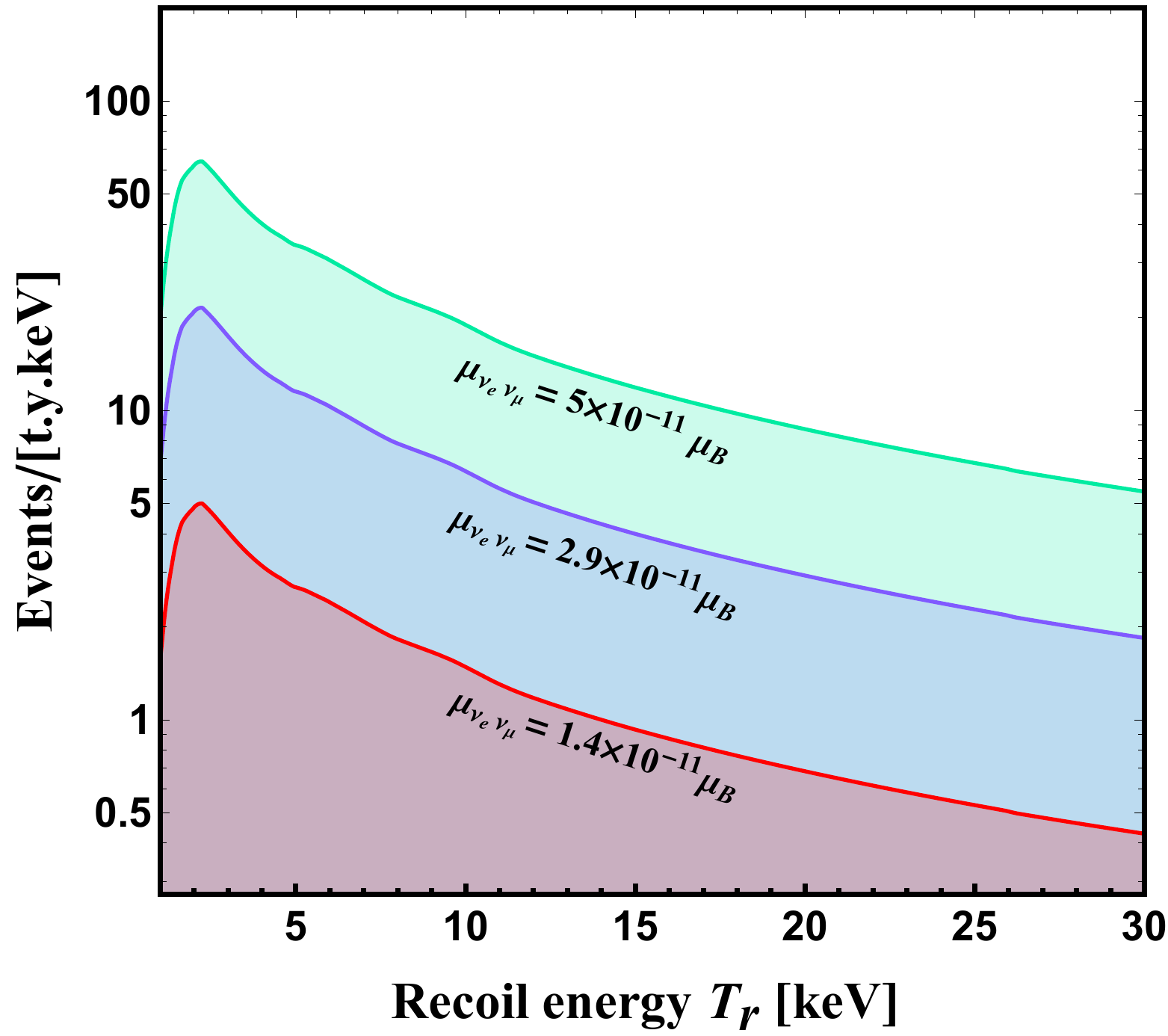}
    \includegraphics[width=.45\textwidth]{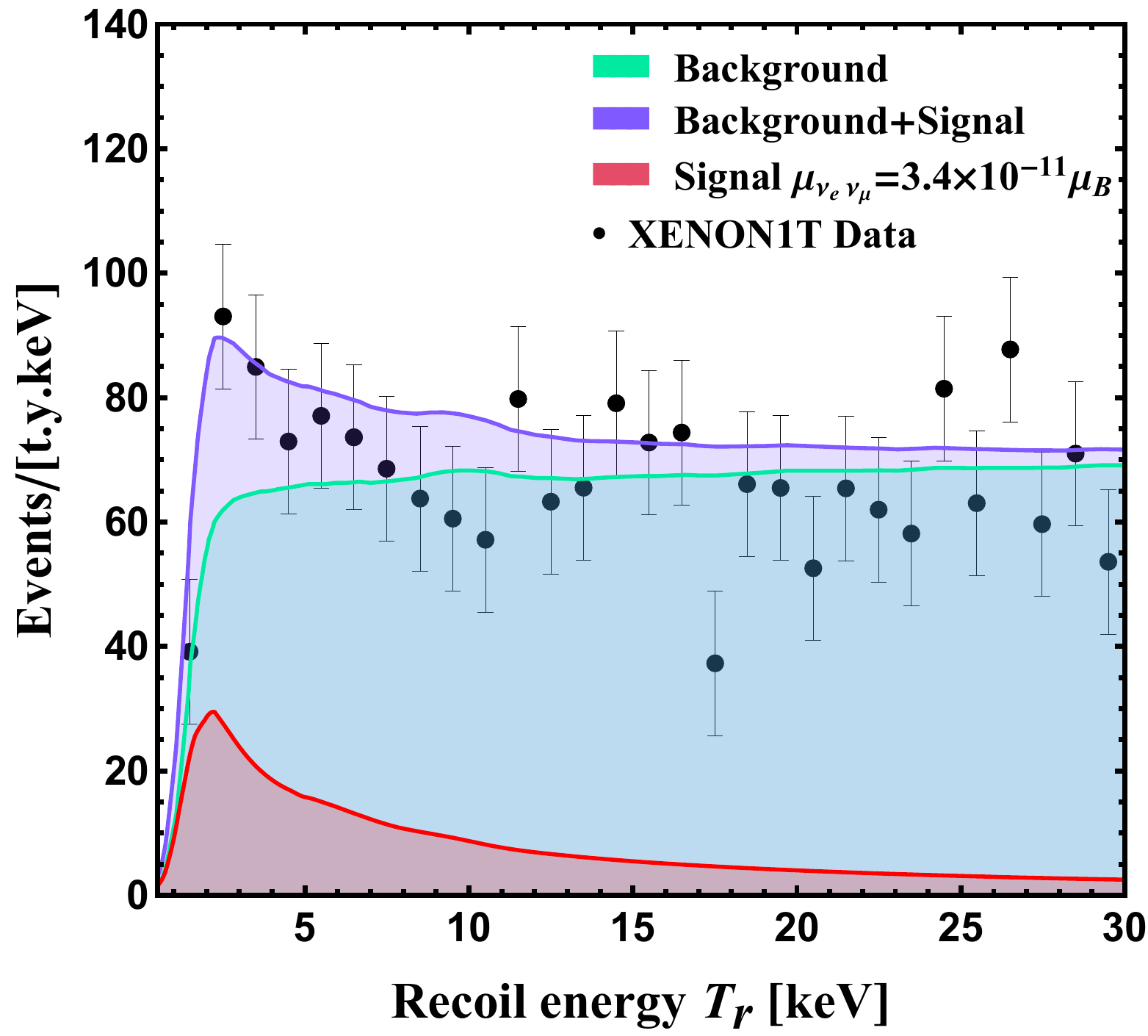}
    \caption{Number of electron recoil events as a function of the recoil energy. Left: green, blue and orange shaded region indicate the signals corresponding to the neutrino magnetic moment $\mu_{\nu_e \nu_\mu}=5\times 10^{-11}\mu_B,\, 2.9\times 10^{-11}\mu_B$ and $1.4\times 10^{-11}\mu_B$ respectively. Right:  Signal spectrum in presence of single component neutrino transition magnetic moment $\mu_{\nu_e \nu_\mu}= 3.4 \times 10^{-11} \mu_B.$ The XENON1T experimental data  \cite{Aprile:2020tmw} and  background spectrum are also shown.    }
    \label{spectrum}
\end{figure}
The excess in electron recoil events observed by XENON1T collaboration \cite{Aprile:2020tmw} may be explained by solar neutrinos which have nonzero magnetic moments. With its low threshold, XENON1T detector is very sensitive to magnetic moments of Dirac neutrinos or to transition moments of Majorana neutrinos, since in either case the neutrino-electron scattering cross-section at low energies will increase. Here we focus on the transition magnetic moment, which is what the models discussed later predict. The differential cross section for the neutrino-electron scattering process $\nu_\alpha e \to \nu_\alpha e$ in the presence of a magnetic moment is given by
\begin{equation}
\left(\frac{d \sigma_{\nu_\alpha e}}{d T}\right)_{t o t}=\left(\frac{d \sigma_{\nu_\alpha e}}{d T}\right)_{S M}+\frac{\pi \alpha^{2}}{m_{e}^{2}}\left(\frac{1}{T}-\frac{1}{E_{\nu}}\right)\left(\frac{\mu_{eff}}{\mu_{B}}\right)^{2}
\end{equation}
where $\mu_{eff}$ is an effective neutrino magnetic moment (defined in Eq. (\ref{eff}) below), ${T}$ is the recoil kinetic energy of the electron and $E_{\nu}$ the  energy of the neutrino. The Standard Model cross section for $\nu_\alpha e \to \nu_\alpha e$ is given by
\begin{equation}
\left(\frac{d \sigma_{\nu_\alpha e}}{d T}\right)_{S M}= \frac{G_{F}^{2} m_{e}}{2 \pi}\left[\left(g_{V}^\alpha +g_{A}^\alpha \right)^{2}+ \left(g_{V}^\alpha-g_{A}^\alpha \right)^{2}\left(1-\frac{T}{E_{\nu}}\right)^{2}+\left(g_{A}^{\alpha^{2}}-g_{V}^{\alpha^2}\right) \frac{m_{e}T}{E_{\nu}^{2}}\right],
\end{equation}
where $\alpha$ represents the neutrino flavor, $m_{e}$ denotes the electron mass and $G_{F}$ is the Fermi constant. The flavor-dependent (since $\nu_e$ undergoes charged current scattering, while $\nu_{\mu,\tau}$ do not) vector and axial vector couplings are given by
\begin{eqnarray}
g_{V}^{e}&=&2 \sin ^{2} \theta_{W}+\frac{1}{2} ; \quad g_{A}^{e}=+\frac{1}{2}\nonumber \\
g_{V}^{\mu, \tau}&=&2 \sin ^{2} \theta_{W}-\frac{1}{2} ; \quad g_{A}^{\mu, \tau}=-\frac{1}{2}~.
\end{eqnarray}

The solar neutrinos flux at low energies is primarily composed of the continuous $pp$-flux and a discrete $^7$Be-flux with values given by \cite{Bahcall:2004mz}
\begin{equation}
    \phi_{pp}=5.94\times 10^{10} \mathrm{cm^{-2}s^{-1}},
\end{equation}
\begin{equation}
    \phi_{^7Be}=4.86\times 10^{9} \mathrm{cm^{-2}s^{-1}}.
\end{equation} 
It is clear that the $pp$ flux is dominant with the $^7$Be
flux an order of magnitude smaller. Flux from $^8$B and other sources are even smaller at low energies.  It is sufficient then to keep only the $pp$ flux in the calculation of electron recoil excess.  $\nu_e$s produced in the solar core oscillate into $\nu_\alpha$ with $\alpha= \mu,\,\tau$, with the flavor transition being adiabatic inside the Sun. 
Since solar neutrinos arriving at earth are a mixture of incoherent states, the effective magnetic moment relevant for the neutrino-electron scattering can be defined as \cite{Grimus:2002vb}:
\begin{equation}
\mu_{e f f}^{2}=\cos^2\theta_{13}\, |\lambda_{12}|^{2}+\left[1-\cos^2\theta_{13} (1-P_{e 1}^{2 \nu})\right] |\lambda_{13}|^{2}+\left(1-\cos^2\theta_{13} P_{e 1}^{2 \nu}\right) |\lambda_{23}|^{2}~.
\label{eff}
\end{equation}
Here $\lambda_{ij} = \mu_{ij} - i d_{ij}$, which contain the transition magnetic and electric dipole moment operators of the physical neutrino states $\nu_{1,2,3}$.  These quantities are related to the transition moments in the flavor basis denoted as $\lambda_{\alpha \beta}$, with $\alpha,\, \beta = e,\, \mu,\, \tau$ via the relation $\tilde{\lambda} = U^T \lambda U$, where $U$ is the PMNS matrix, with $\tilde{\lambda}$ denoting  $\lambda_{ij}$ in the mass eigenstate basis, and $\lambda$ denoting  $\lambda_{\alpha\beta}$ in the flavor basis.  
In Eq. (\ref{eff}) $P_{e 1}^{2 \nu}$ denotes the probability of observing the mass eigenstate $\nu_1$ at the scattering point for an initial electron flavor in the  two-neutrino oscillation scenario. It is clear from Eq. (\ref{eff}) that CP violating phases of the PMNS matrix do not affect $\mu_{eff}^2$.  

We shall be interested in a scenario where only the $\mu_{\nu_e\nu_\mu}$ component of the magnetic moment matrix, expressed in the flavor basis, is nonzero.  When converting this into the mass eigenbasis so that Eq. (\ref{eff}) can be used, all the neutrino oscillation parameters come into play, including the Dirac CP phase $\delta$.  We use central values of the oscillation parameters given in Ref. \cite{Esteban:2018azc}, viz., $\{\sin^2\theta_{12} = 0.310,\, \sin^2\theta_{23} = 0.580,\, \sin^2\theta_{13} =0.0224,\, \delta_{CP} = 215^\circ\}$.  For the effective 2-neutrino oscillation probability, we use the best fit value $P_{e 1}^{2 \nu}=0.667$ \cite{Miranda:2020kwy}. With these, we can express the effective 
 neutrino magnetic moment in terms of single component transition neutrino magnetic moment (in flavor basis) as:
\begin{equation}
    \mu_{e f f}^{2}=0.72\, \mu_{\nu_e\nu_\mu}^2.
    \label{number}
\end{equation}
We shall use this value in our numerical analysis.\footnote{If we use the coefficient on the right hand side of  Eq. (\ref{number}) to be 1, we have verified that the XENON1T \cite{Aprile:2020tmw} analysis can be reproduced.} The transition dipole moments in the mass basis, for this choice of parameters, are found to be
\begin{equation}
   \left( |\lambda_{12}|,\, |\lambda_{13}|,\, |\lambda_{23}|\right) = \left( 0.65,\, 0.59,\, 0.49 \right) \times |\mu_{\nu_e \nu_\mu}|~.
\end{equation}

Here we have focused on the case of Majorana neutrinos, for which only the transition magnetic moments are nonzero due to CPT-conservation. If neutrinos are Dirac particles, all elements of the magnetic moment matrix could have nonzero values. 

In order to compute XENON1T signal prediction and analyze the recoiled electron spectrum  for  a single component transition magnetic moment $\mu_{\nu_e \nu_\mu}$,  one can define the differential event rate in terms of the reconstructed recoiled energy ($T$) as
\begin{equation}\frac{d N}{d T_r}=n_{e} \times \int_{E_{\nu}^{min}}^{E_{\nu}^{max}} d E_{\nu} \int_{T^{t h}}^{T^{max}} d T\left(\frac{d \sigma_{\nu_{e} e}}{d T} P_{e e}+\frac{d \sigma_{\nu_{\mu } e}}{d T} P_{e \mu }\right) \times \frac{d \phi}{d E_{\nu}} \times \epsilon\left(T \right) \times \mathcal{G}\left(T, T_{r}\right)\end{equation}
where   $d \phi / d E_{\nu}$ denotes the solar neutrino flux spectrum \cite{Bahcall:2004mz}, $\epsilon\left(T\right)$ indicates the detector efficiency  \cite{Aprile:2020tmw, Aprile:2020yad}, $n_e$ is the number of target electrons in fiducial volume of one ton Xenon  \cite{Aprile:2020tmw} and  $\mathcal{G}\left(T, T_{r}\right)$ represents a normalized Gaussian smearing function in order to account for the detector finite energy resolution \cite{Aprile:2020tmw,Aprile:2020yad}. The detector threshold and the maximum recoil energy are respectively given by $T^{t h}=1 $ {keV}  and $T^{max}=30 $ keV, while the other integration limits are $ E_{\nu}^{min}=\left(T+\sqrt{2 m_{e} T+T^{2}}\right) / 2 \text { and } E_{\nu}^{max}=420$ keV. 

By folding the expected solar neutrino flux \cite{Bahcall:2004mz}  and imposing a step-function approximation to account for the electron binding energies, we analyze the recoiled energy spectrum for different values of neutrino transition magnetic moment 
$\mu_{\nu_e \nu_\mu} = \{1.4 \times 10^{-11},\, 2.9 \times 10^{-11}, \,5 \times 10^{-11}\} \mu_{B}$ in Fig.~\ref{spectrum}. For this analysis, we adopt the background model spectrum from Ref.  \cite{Aprile:2020tmw}. The preferred values of neutrino transition magnetic moment  $\mu_{\nu_e \nu_\mu}$  for the excess observed at XENON1T experiment \cite{Aprile:2020tmw} at  $90 \%$ confidence interval corresponds to $\mu_{\nu_e \nu_\mu} \in(1.65,3.42) \times 10^{-11} \mu_{B}.$
In the right panel of Fig.~\ref{spectrum} we show results of our analysis of the signal and background spectrum where we also compare ths with the observed data \cite{Aprile:2020tmw}. Red shaded zone indicates the signal spectrum only corresponding to $\mu_{\nu_e \nu_\mu}=3.4\times 10^{-11}\mu_B$. The green shaded region indicates the background spectra and purple shaded zone shows the expected combined spectrum for signal and background. One sees that owing to the presence of sizable neutrino magnetic moment, and the resulting $1/T$ enhancement in the cross section, the signal spectrum gives a good fit to the observed data in the electron recoil energy range between $(1 - 7)$ keV peaking around $2.5$ keV.  This shows the consistency of a single component transition magnetic moment interpretation of the Xenon data.

\subsection{Experimental searches for neutrino magnetic moments}

The quest for measuring a possible magnetic moment of the neutrino was begun even before the discovery of the neutrino. Cowan, Reines and Harrison set an upper limit of $\mu_\nu < 10^{-7} \mu_B$ in the process of measuring background for a free neutrino search experiment \cite{Cowan:1954pq} with reactor antineutrinos.  This limit is obtained by studying $\overline{\nu}_e e$ elastic scattering process and observing a possible excess in the electron recoil events.  This Cowan-Reines-Harrison limit was subsequently improved by several orders of magnitude by a variety of reactor antineutrino experiments.  KRASNOYARSK reactor experiment obtained a limit of $\mu_\nu < 2.7 \times 10^{-10} \mu_B$ \cite{Vidyakin:1992nf}, with subsequent improvements by ROVNO ($\mu_\nu < 1.9 \times 10^{-10}\mu_B$) \cite{Derbin:1993wy}, MUNU ($\mu_\nu < 1.2 \times 10^{-10} \mu_B$) \cite{Daraktchieva:2005kn} and TEXONO ($\mu_\nu < 2 \times 10^{-10}\mu_B$) \cite{Deniz:2009mu}. 
The GEMMA collaboration reports a more stringent limit on $\overline{\nu}_e$ magnetic moment of $\mu_\nu < 2.9 \times 10^{-11} \mu_B$ \cite{Beda:2012zz}. 
These limits apply specifically to either a Dirac magnetic moment or a Majorana transition magnetic moment of $\overline{\nu}_e$.  

Accelerator based experiments have also searched for neutrino magnetic moments via low energy $\nu_e,\,\nu_\mu$ and $\overline{\nu}_\mu$ scattering off electrons.  By studying $\nu_e-e$ scattering, a bound on an effective magnetic moment has been obtained at LAPMF which translates into a muon neutrino magnetic moment limit of $\mu_{\nu_\mu} < 7.4 \times 10^{-10} \mu_B$ \cite{Allen:1992qe}. LSND experiment has obtained a limit of $\mu_{\nu_\mu} < 6.4 \times 10^{-10} \mu_B$, also by studying $\nu_e-e$ scattering.  

The Borexino experiment has studied the shape of the electron recoil spectrum from solar neutrino interactions and found no significant deviations from expectations.  A limit on an effective neutrino magnetic moment $\mu^{\rm eff}_\nu < 2.8 \times 10^{-11} \mu_B$ was obtained \cite{Borexino:2017fbd}.  When interpreted as a single component Majorana neutrino transition magnetic moment, this would translate into $\mu_{\nu_e\nu_\mu} < 3.29 \times 10^{-11}\mu_B$.  This limit, which is more directly related to the XENON1T excess, is consistent with the needed value to explain the excess. For a global fit including all the experimental limits on neutrino magnetic moments, see Ref. \cite{Canas:2015yoa,Miranda:2020kwy} and also Ref. \cite{Khan:2020vaf}. 

\subsection{Limits on $\boldmath{\mu_\nu}$ from astrophysics and cosmology}

Evolution of stars can provide indirect constraints on
the magnetic moments of either Dirac  or Majorana neutrinos.  Photons in the plasma of stellar environments can decay either into $\nu \overline{\nu}$ for the case of Dirac neutrinos or into $\nu_\alpha \nu_\beta$ for the case of Majorana neutrinos \cite{Bernstein:1963qh,Raffelt:1999tx}.  Such decays are kinematically allowed in a plasma since the photon acquires a mass.  If such decays occur too rapidly, that would drain energy of the star, in conflict with standard stellar evolution models which appear to be on strong footing.  Limits on $\mu_\nu$ have been derived by requiring the energy loss in such decays to be not more than via standard processes.  The best limit on $\mu_\nu$ from this argument arises from red giant branch of globular clusters, resulting in a limit of $\mu_\nu < 4.5 \times 10^{-12} \mu_B$ \cite{Viaux:2013lha}.  Validity of this limit would make the neutrino magnetic moment interpretation of the XENON1T excess questionable.  We note that these indirect constraints from  astrophysics may be evaded if the plasmon decay to neutrinos is kinematically forbidden.  As we show in Sec. \ref{SEC-08}, this can indeed be achieved by invoking interactions of the neutrino with a light scalar. The neutrino will then acquire a medium-dependent mass greater than the plasmon mass, while being consistent with other observations, and thus forbidding plasmon decays. There are also cosmological limits on $\mu_\nu$ arising from big bang nucleosynthesis. However, these limits are less severe, of order $10^{-10}\mu_B$ \cite{Vassh:2015yza}.

We now turn to theoretical interpretation of the suggested transition magnetic moment $\mu_{\nu_e \nu_\mu}$.

\section{New symmetries and a large neutrino magnetic moment}\label{SEC-03}
In this section we  recall  theoretical expectations for neutrino magnetic moment and revive symmetry based mechanisms to generate sizable $\mu_\nu$.

\subsection{Neutrino magnetic moment in the Standard Model and beyond}

The magnetic moment and mass operators for the neutrino have the same chiral structure, which for a Dirac neutrino has the form:
\begin{equation}
{\cal L} \supset \mu_\nu \overline{\nu}_L \sigma_{\mu\nu} \nu_R F^{\mu\nu} + m_\nu \overline{\nu}_L \nu_R + {\rm H.c.}
\end{equation}
As a result, $\mu_\nu$ typically becomes proportional to $m_\nu$.  For example, in the Standard Model when right-handed neutrinos are introduced so that the neutrino has a small Dirac mass, its magnetic moment is given by \cite{Fujikawa:1980yx}
\begin{equation}
    \mu_\nu = \frac{ e G_F m_\nu}{8 \sqrt{2}\pi^2} = 3 \times 10^{-20} \mu_B\, \left(\frac{m_\nu}{0.1~{\rm eV}}\right)~.
    \label{dirac}
\end{equation}
If neutrinos are Majorana particles, their transition magnetic moments resulting from Standard Model interactions is given by \cite{Pal:1981rm}
\begin{equation}
    \mu_{ij} = -\frac{3eG_F}{32\sqrt{2}\pi^2}(m_i \pm m_j)\sum_{\ell = e, \mu,\tau}U_{\ell i}^* U_{\ell j} \frac{m_\ell^2}{m_W^2},
\end{equation}
where $m_i$ stands for mass of neutrino $i$, $m_\ell$ is the charged lepton mass, and $U_{\ell i}$ denotes the PMNS matrix element.  The resulting transition magnetic moment is even smaller than the value given in Eq. (\ref{dirac}), at most of order $10^{-23} \mu_B$. Clearly, these values are well below the sensitivity of current experiments.

Nonstandard interactions of the neutrinos can lead to enhanced magnetic moments, esepcially when the new physics lies near the TeV scale. For example, in left-right symmetric models, the right-handed neutrino couples to a $W_R^\pm$ gauge boson, which also has mixing with the $W$ boson. For the case of a Dirac neutrino the magnetic moment now becomes proportional to the charged lepton mass, rather than $m_\nu$, and is given by
\begin{equation}
    \mu_\nu \simeq \frac{G_F\,m_\ell}{2 \sqrt{2} \pi^2}  \sin2\xi,
\end{equation}
where $\xi$ is the mixing angle between $W_R^\pm$ and $W^\pm$, which is of order $(M_W^2/M_{W_R}^2)$.  This mixing angle is constrained by muon decay asymmetry parameters \cite{Bayes:2011zza}, as well as by $b \rightarrow s\gamma$ decay rate \cite{Babu:1993hx}, leading to a limit $\mu_{\nu_e} < 10^{-14} \mu_B$ \cite{Giunti:2014ixa}.   While significantly enhanced compared to the Standard Model value of Eq. (\ref{dirac}), this is still well below experimental sensitivity.

In supersymmetric extensions of the Standard Model, lepton number may be violated by $R$-parity breaking interactions.  In such contexts, without relying on additional symmetries, the neutrino transition magnetic moment will be of the order $\mu_\nu \sim \lambda^{\prime 2}/(16 \pi^2) m_\ell^2 A_\ell /M_{\tilde{\ell}}^4$, where
$\lambda^{\prime}$ is an $R$-parity breaking coupling, $A_\ell$ is the SUSY breaking trilinear coupling, and
$M_{\tilde{\ell}}$ is the slepton mass.  Imposing experimental constraints on the SUSY parameters, this would yield a value of $\mu_\nu$ at most about $10^{-13} \mu_B$, which is too small to be relevant for XENON1T. Transition magnetic moments can be larger in presence of new vector-like leptons \cite{Kim:1976gk}. 

It is possible to induce $\mu_\nu \sim {\rm few} \times 10^{-11}\mu_B$ via charged scalar loops, which are less constrained by other processes.  An $SU(2)_L$ singlet charged scalar $\eta^+$ can induce significant $\mu_\nu$ for a Dirac neutrino or to a Majorana neutrino of the desired order to explain the XENON1T anomaly \cite{Fukugita:1987ti,Babu:1987be}.  However, even in this case, the neutrino mass -- magnetic moment problem shown in Eq. (\ref{order}) would prevail.  While $\mu_\nu$ can be large as desired, $m_\nu$ will become unacceptably large, unless it is strongly fine-tuned to about one part in $10^6$.

\subsection{\boldmath$SU(2)_H$ symmetry for enhanced neutrino magnetic moment}

While the neutrino mass operator and the magnetic moment operator both are chirality flipping, there is one important difference in their Lorentz structures. The mass operator, being a Lorentz scalar, is symmetric, while the magnetic moment, being a Lorentz tensor operator is antisymmetric in the two fermion fields.  Voloshin suggested to exploit this property to suppress neutrino mass while enhancing its magnetic moment \cite{Voloshin:1987qy}. He proposed a new $SU(2)_\nu$ symmetry that transforms $\nu$ into $\nu^c$, the left-handed antiparticle of the right-handed neutrino. A neutrino mass term, being symmetric under the exchange of $\nu$ and $\nu^c$, would then be forbidden by the $SU(2)_\nu$ symmetry, which requires such an invariant to be antisymmetric, since a singlet made out of two $SU(2)_\nu$ doublets is in the antisymmetric combination.  On the other hand, the magnetic moment operator, $\nu^T C \sigma_{\mu \nu}\nu^cF^{\mu \nu}$ is antisymmetric under  $\nu \leftrightarrow \nu^c$ interchange, and thus is allowed in the $SU(2)_\nu$ symmetric limit.

As it turns out, since the $\nu^c$ field does not feel weak charged current or neutral current interactions, the $SU(2)_\nu$ symmetry operating on $(\nu,\, \nu^c)$ fields is not easy to implement \cite{Barbieri:1988fh}. It was suggested in Ref. \cite{Babu:1989wn} that a horizontal $SU(2)_H$ symmetry acting on the electron and the muon families can serve the same purpose, which is easier to implement as such a symmetry commutes with the weak interactions.  This would lead to a transition magnetic moments for Majorana neutrinos. Models based on such $SU(2)_H$ symmetries were built, which we shall revive and simplify in the next section.  The main point of the $SU(2)_H$ symmetry is that the neutrino transition magnetic moment interaction given by
\begin{eqnarray}
{\cal L}_{\rm mag.} = (\nu_e^T ~~ \nu_\mu^T) C^{-1} \sigma_{\mu \nu}\left( \begin{matrix} 0 & 1 \\ -1 & 0 \end{matrix}\right)\left( \begin{matrix} \nu_e \\ \nu_\mu \end{matrix}\right) F^{\mu \nu},
\end{eqnarray}
where $C$ is the charge conjugation matrix, is invariant under any $SU(2)_H$ transformations as $U^T i\tau_2 U = i \tau_2$ for any $2 \times 2$ unitary matrix $U$ that rotates $\nu_e$ and $\nu_\mu$.  On the other hand, the Majorana neutrino mass term given by
\begin{eqnarray}
{\cal L}_{\rm mass} = (\nu_e^T ~~ \nu_\mu^T) C^{-1} \left( \begin{matrix} 0 & 1 \\ 1 & 0 \end{matrix}\right)\left( \begin{matrix} \nu_e \\ \nu_\mu \end{matrix}\right)
\end{eqnarray}
is not invariant under a unitary rotation involving ($\nu_e,\, \nu_\mu)$. Thus, in the $SU(2)_H$ symmetric limit, neutrino mass is forbidden, while a transition magnetic moment $\mu_{\nu_e \nu_\mu}$ is permitted.  Explicit realization of this idea was given in Ref. \cite{Babu:1989wn,Babu:1990wv}. %It was also realized that the full $SU(2)_H$ symmetry may not be essential, a non-Abelian discrete subgroup of it with certain properties would suffice \cite{Ecker:1989ph,Babu:1989px,Chang:1990uga,Leurer:1989hx,Choudhury:1989pw}. 

It has been also realized that the full $SU(2)_H$ symmetry is not essential to realize a large $\mu_\nu$ with a suppressed mass, a non-Abelian subgroup of $SU(2)_H$ would suffice \cite{Ecker:1989ph,Babu:1989px,Chang:1990uga,Leurer:1989hx,Choudhury:1989pw}.  To see this, note that an $i\tau_2$ rotation would prevent an off-diagonal neutrino mass, while an $i\tau_3$ rotation would prevent any diagonal masses.  The magnetic moment operator is invariant under both rotations.  Since $i\tau_2$ rotation does not commute with $i\tau_3$ rotation, the full symmetry group should contain at least eight elements: $\{\pm 1,\,\pm i\tau_1,\,\pm i \tau_2,\,\pm i\tau_3\}$.  The quaternion group of order 8 is an example of such a symmetry.  For a review of these developments see Ref. \cite{Babu:1991zh}; for a recent update see Ref. \cite{Lindner:2017uvt}.  

In the next section we shall present a model based on {\it approximate} $SU(2)_H$ symmetry to induce a large $\mu_{\nu_e\nu_\mu}$.  Since the symmetry is only approximate, there is no significant difference between models based on $SU(2)_H$ or one of its non-Abelian subgroups.  By requiring only an approximate $SU(2)_H$, as opposed to exact symmetry of Ref. \cite{Babu:1989wn,Babu:1990wv}, the model becomes simpler. It should be noted that the mass splitting between the electron and the muon breaks the approximate $SU(2)_H$ symmetry.  If all violations of $SU(2)_H$ are of the order of $(m_\mu^2 - m_e^2)/M^2$, where $M$ is a heavy mass scale of order 100 GeV or more, then the naive estimate of Eq. (\ref{order}) would be modified to 
\begin{equation}
m_\nu \sim \frac{ \mu_\nu} {\mu_B} \frac{M^2} {2 m_e} \frac{(m_\mu^2 - m_e^2)}{M^2} =  \frac{ \mu_\nu} {\mu_B}   \frac{(m_\mu^2 - m_e^2)}{2 m_e}~.
\end{equation}
This estimate will give $m_\nu \sim 0.1$ eV for $\mu_\nu \sim 10^{-11} \mu_B$, which is just about acceptable.  In fact, we shall see that there is in addition, a loop suppression factor, which would make $m_\nu$ related to the magnetic moment operator smaller by another two orders of magnitude.

\subsection{Large magnetic moment from spin symmetry}

A somewhat independent mechanism is known for generating enhanced $\mu_\nu$ with a suppressed $m_\nu$.  This relies on a spin symmetry argument.  In renormalizable gauge theories there are no direct couplings of the type $\gamma W^+ S^-$ where $S^-$ is a charged scalar field.  However, such a coupling could be generated via loops.  Barr, Friere and Zee \cite{Barr:1990um} used this induced vertex to construct models of large $\mu_\nu$.  At the two loop level, this vertex will contribute to $\mu_\nu$.  As for its contribution to $m_\nu$, it is well known that for transversely polarized vector bosons, the transition from spin 1 to spin 0 cannot occur. Only the longitudianl mode, the Goldstone mode, would contribute to such transitions.  This implies that in the two loop diagram utilizing the $\gamma W^+ S^-$ for generating $\mu_\nu$, if the photon line is removed, only the longitudinal $W^\pm$ bosons will contribute, leading to a suppression factor of $m_\ell^2/m_W^2$ in the neutrino mass, compared to the naive estimate of Eq. (\ref{order}).  

This idea of utilizing spin symmetry has a simple realization in the popular Zee model of neutrino masses \cite{Zee:1980ai}, as was shown in Ref. \cite{Babu:1992vq}.  We have investigated the current status of neutrino magnetic moment in this class of models.  We found that while these models typically induce large $\mu_\nu$, after taking account of low energy constraints as well as LHC constraints on new particles, the maximun $\mu_\nu$ that can be generated here is about an order of magnitude smaller than the value needed to explain XENON1T anomaly.

\subsection{Dirac vs Majorana neutrino magnetic moments}

It has been argued, based on effective field theory (EFT) calculations,  that Dirac neutrino magnetic moments exceeding about $10^{-15}\mu_B$ would not be natural, as that would induce at higher loops unacceptably large neutrino masses \cite{Bell:2005kz}. For Majorana neutrinos, the transition magnetic moments are allowed to be much larger from EFT naturalness arguments \cite{Davidson:2005cs,Bell:2006wi}.  For example, $\mu_{\nu_e\nu_\mu} \sim 10^{-7}\mu_B$ would be allowed by EFT, if the new physics scale is around a TeV. In the case of a Dirac neutrino magnetic moment, weak interaction corrections to the neutrino mass arising from the magnetic moment operator are excessive, while such corrections are proportional to charged lepton mass differences and small in EFT for the case of Majorana neutrinos. Thus, if neutrino magnetic moments are measured at the level of current experimental sensitivity, it is very likely that neutrinos are Majorana particles.  The $SU(2)_H$ symmetry based models, as well as the spin symmetry based models, fit well within this categorization.

%%%%%%%%%%%%%%%%%%%%%%%%%%%%%%%%%%%%%%%%%%%%%%%
%%%%%%%%%%%%%%%%%%%%%%%%%%%%%%%%%%%%%%%%%%%%%%%
\section{ \boldmath${SU(2)_H}$ model for large neutrino magnetic moment }\label{SEC-04}

In this section we present a simplified model for large transition magnetic moment $\mu_{\nu_e\nu_\mu}$ based on an {\it approximate} $SU(2)_H$ horizontal symmetry acting on the electron and the muon families.  A full $SU(2)_H$ symmetric model was presented in Ref. \cite{Babu:1989wn,Babu:1990wv}, which is our starting point. Our simplification is that the symmetry is only approximate, broken explicitly by electron and muon masses.  Fewer new particles would then suffice to complete the model.  The explicit breaking of $SU(2)_H$ by the lepton masses is analogous to chiral symmetry breaking in the strong interaction sector by masses of the light quarks.  Such breaking will have to be included in the neutrino sector as well.  We have computed the one-loop corrections to the neutrino mass from these explicit breaking terms and found them to small enough so as to not upset the large magnetic moment solution.

The only violation of $SU(2)_H$ acting on the electron and muon fields arises from their unequal masses.  This mass splitting, normalized to the weak scale, is indeed a small parameter: $(m_\mu^2-m_e^2)/m_W^2
= 1.7 \times 10^{-6}$.  Violation of $SU(2)_H$ symmetry in the neutrino masses can be of this order, which from Eq. (\ref{order}) suggests that large $\mu_{\nu_e\nu_\mu}$ can be realized without inducing large $m_\nu$.  In fact, the effect of the $SU(2)_H$ breaking parameter $(m_\mu^2-m_e^2)/m_W^2$ in the neutrino sector will be accompanied by a loop suppression factor of order $10^{-2}$, which would make $m_\nu$ even smaller.

Our model is a simple extension of the Zee model \cite{Zee:1980ai} of neutrino mass that accommodates an $SU(2)_H$ symmetry.  The Zee model is one of the simplest models of neutrino mass generation with new scalars possibly having masses in the TeV scale.  A sizable neutrino transition magnetic moment requires such particles, along with violation of lepton number.

The gauge symmetry of the model is $SU(3)_c \times SU(2)_L \times U(1)_Y$, with no new fermions added to the Standard Model.  In addition, there is an approximate $SU(2)_H$ symmetry. Leptons of the Standard Model transform under $SU(2)_L \times U(1)_Y \times SU(2)_H$ as follows:
\begin{eqnarray}
\psi_L &=& \left(\begin{matrix} \nu_e & \nu_\mu \\ e & \mu \end{matrix} \right)_L~~~~~~(2, -\frac{1}{2}, 2) \nonumber \\
\psi_R &=& ~~(e~~~~~~\mu)_R ~~~~~~~(1,-1,2) \nonumber \\
\psi_{3L} &=&~~~ \left( \begin{matrix}\nu_\tau \\ \tau  \end{matrix} \right) ~~~~~~~~~~~(2,-\frac{1}{2}, 1)\nonumber\\ 
&~&~~~~~~\tau_R ~~~~~~~~~~~~~(1,-1,1)~.
\end{eqnarray} 
Here $SU(2)_H$ acts horizontally, while $SU(2)_L$ acts vertically.  
The first two families of leptons form a doublet of $SU(2)_{H}$ while the $\tau$ family is a singlet. All quark fields are assumed to be $SU(2)_H$ singlets.  

The Higgs sector of the model consists of the following multiplets:
\begin{eqnarray}
\phi_{S}=\left(\begin{array}{c}
\phi_{S}^{+} \nonumber \\
\phi_{S}^{0}
\end{array}\right) &   \quad \quad \quad \left(2, \frac{1}{2}, 1\right) \nonumber \\
\Phi=\left(\begin{array}{cc}
\phi_{1}^{+} & \phi_{2}^{+} \nonumber \\
\phi_{1}^{0} & \phi_{2}^{0}
\end{array}\right) & \quad \quad \quad  \left(2, \frac{1}{2}, 2\right) \\
\eta=(\eta_{1}^{+} \quad \eta_{2}^{+}) &   \quad \quad \quad (1,1,2)~.
\end{eqnarray}
The $\phi_S$ filed is the Standard Model Higgs doublet, which has its usual Yukawa couplings with the quarks.  The $\phi_S$ field is also responsible for electroweak symmetry breaking.  The vacuum expectation values (VEV) of $\phi_S^0$ is denoted as $\left\langle \phi_S^0\right\rangle = v/\sqrt{2}$ where $v \simeq 246$ GeV. The $\phi$ fields are assumed to acquire no VEVs. This is a consistent assumption, which is valid  even after the explicit breaking of $SU(2)_H$ symmetry.

Under $SU(2)_L \times SU(2)_H$, the transformation of various fields is as follows:
\begin{eqnarray}
       ~&~ \psi_L \rightarrow U_L \psi_L U_H^T,~~~\psi_{\tau L} \rightarrow U_L \psi_{\tau L},~~~\psi_R \rightarrow \psi_RU_H^T \nonumber \\
      ~&~  \phi_S \rightarrow U_L \phi_S,~~~\Phi \rightarrow U_L \Phi U_H^T,~~~\eta \rightarrow \eta U_H^T~.
\end{eqnarray}
Here $U_L$ and $U_H$ are unitary matrices associated with $SU(2)_L$ and $SU(2)_H$ transformations.  The Yukawa Lagrangian in the lepton sector that is invariant under the gauge symmetry as well as $SU(2)_H$ is then
\begin{eqnarray}
\mathcal{L}_{\rm Yuk} &=& h_{1} \operatorname{Tr}\left(\bar{\psi}_{L} \phi_{S} \psi_{R}\right)+h_{2} \bar{\psi}_{3 L} \phi_{S} \tau_{R}+h_{3} \bar{\psi}_{3 L} \Phi i \tau_{2} \psi_{R}^{T} \nonumber \\
&~& f \eta \tau_{2} \psi_{L}^{T} \tau_{2} C \psi_{3 L}+f^{\prime} \operatorname{Tr}\left(\bar{\psi}_{L} \Phi\right) \tau_{R} + H.c.
\label{Yuk1}
\end{eqnarray}
Expanded in component form, this reads as:
\begin{eqnarray}
\mathcal{L}_{\rm Yuk} &=& h_{1}\left[(\bar{\nu}_{e} e_{R} \phi_{S}^{+}+\bar{e}_{L} e_{R} \phi_{S}^{0})+(\bar{\nu}_{\mu} \mu_{R} \phi_{S}^{+}+\bar{\mu}_{L} \mu_{R} \phi_{S}^0)\right] 
+h_{2}\left[\bar{\nu}_{\tau} \tau_{R} \phi_{S}^{+}+\bar{\tau}_{L} \tau_{R} \phi_{S}^{0}\right] \nonumber\\
&+& h_{3}\left[-(\bar{\nu}_{\tau} e_{R} \phi_{2}^{+} +\bar{\tau}_{L} e_{R} \phi_{2}^{0}) + (\bar{\nu}_{\tau} \mu_{R} \phi_{1}^{+}+\bar{\tau}_{L} \mu_{R} \phi_{1}^{0})\right]\nonumber\\
&+& f\left[\left(\nu_{e}^T C \tau_{L} - e_{L}^{T} C \nu_\tau\right) \eta_{2}^{+}-\left(\nu_{\mu}^{T} C \tau_{L}-\mu_{L}^{T} C \nu_{\tau} \right) \eta_{1}^{+}\right]\nonumber\\
&+& f^{\prime}\left[(\bar{\nu}_{e} \tau_{R} \phi_{1}^{+}+\bar{e}_{L} \tau_{R} \phi_{1}^{0})+(\bar{\nu}_{\mu}  \tau_{R} \phi_2^{+}+\bar{\mu}_{L} \tau_{R} {\phi_2^{0}})\right] + H.c. \label{yukawasu2}
\end{eqnarray}
It becomes clear that 
the $h_{1}$ term gives equal mass for the electron and the muon once $\left\langle \phi_S^0 \right \rangle = v/\sqrt{2}$ develops. The $h_{2}$ term generates a mass for the $\tau$ lepton. If $h_{3}=0, \tau$ lepton number would be a good symmetry of the Lagrangian. The $h_3$ term induces a nonzero $\nu_\tau$ mass in conjunction with the $f$ term, which is allowed in the limit of exact $SU(2)_H$. The terms $f$ and $f^{\prime}$ are crucial for the generation of the neutrino transition magnetic moment. We shall introduce explicit breaking of the $SU(2)_H$ symmetry, so that the relation $m_e = m_\mu$ which follows from Eq. (\ref{yukawasu2}) can be corrected.  Since this breaking is small, we first discuss the model in the $SU(2)_H$ symmetric limit.

The scalar potential contains, among other terms, the following  terms:\footnote{The full scalar potential has been analyzed in Ref. \cite{Babu:1990wv}}
\begin{equation}
V \supset m_\eta^2(|\eta_1|^2+|\eta_2|^2) + m_{\phi^+}^2 ( |\phi_1^+|^2 + |\phi_2^+|^2) + m_{\phi^0}^2 (|\phi_1^0|^2 + |\phi_2^0|^2) + \{\mu \eta \Phi^{\dagger} i \tau_{2} \phi_{S}^{*} + H.c.\}
\label{potential}
\end{equation}
Here $m_\eta^2$  includes the bare mass term as well as a contribution from the quartic coupling $\lambda_\eta \eta^\dagger \eta \Phi_S^\dagger \Phi_S$, with the VEV of $\phi_S^0$ inserted.  Similarly, $m_{\phi^+}^2$ includes the bare mass of $\phi$ as well as the contribution from the quartic coupling $\lambda_\phi \phi_S^\dagger \phi_S{\rm Tr}(\Phi^\dagger \Phi)$. The mass of $\phi^0$ is split from that of $\phi^+$ through the interaction term ${\rm Tr}|\Phi^\dagger \phi_S|^2$.  All terms in Eq. (\ref{potential}) respect $SU(2)_H$ symmetry.  
In component form the cubic coupling reads as:
\begin{equation}
V^{(3)}=\mu\left[\overline{\phi_S^0} \left(\eta_{1}^{+} \phi_{1}^{-}+\eta_{2}^{+} \phi_{2}^{-}\right)-\phi_{S}^{-}\left(\eta_{1}^{+} \bar{\phi}_{1}^{0}+\eta_{2}^{+} \bar{\phi}_{2}^{0}\right)\right] + H.c.
\label{cubic1}
\end{equation}
Once the VEV of $\phi_S^0$ is inserted, this term would lead to the mixing of $\eta_1^+$ with $\phi_1^+$ and $\eta_2^+$ with $\phi_2^+$.  These mass matrices are identical, owing to the unbroken $SU(2)_H$ and are given by
\begin{eqnarray}
M^2_{\eta_1^+-\phi_1^+} =   M^2_{\eta_2^+-\phi_2^+} = \left(\begin{matrix}m_\eta^2 & \frac{\mu v}{\sqrt{2}} \\ \frac{\mu^* v}{\sqrt{2}} & m_{\phi^+}^2    \end{matrix} \right)~.      
\end{eqnarray}
The two mass eigenstates will be denoted as $h_i^+$ and $H_i^+$ wtih $i = 1,2$ and  their masses will be denoted as $(m_{h^+}^2,\, m_{H^+}^2)$. These states are related to the original states via the relations
\begin{equation}
    h_i^+ = \cos\alpha\, \eta_i^+ + \sin\alpha\, \phi_i^+,~~~~~~H_i^+ = -\sin\alpha \,\eta_i^+ + \cos\alpha\, \phi_i^+~.
\end{equation}
where the mixing angle is given by
\begin{equation}
    \tan2\alpha = \frac{\sqrt{2} \mu v}{m_\eta^2 - m_{\phi^+}^2}~.
\end{equation}

\subsection{Neutrino transition magnetic moment}

\begin{figure}[h!]
    \centering
      \includegraphics[width=\textwidth]{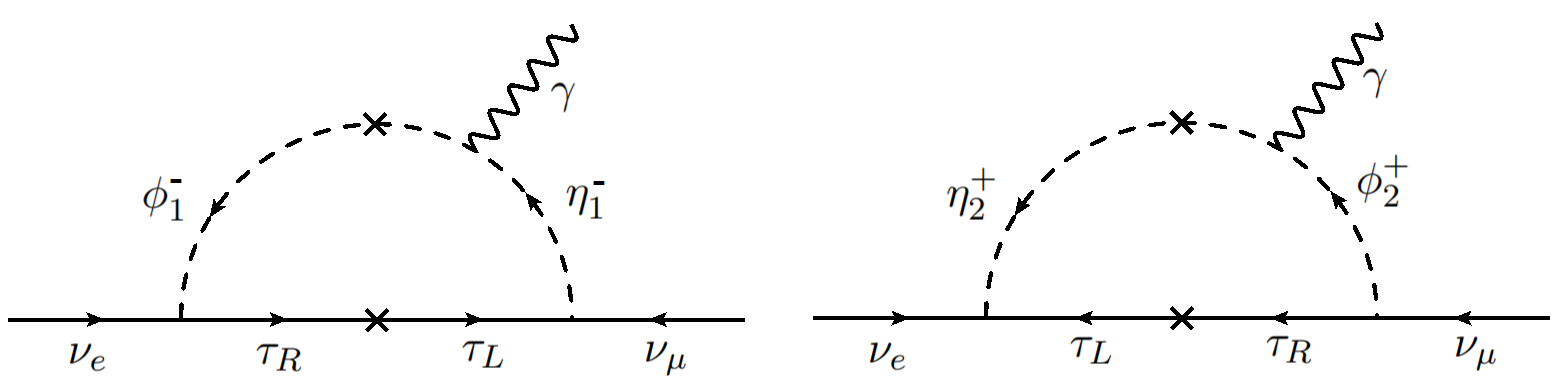}
    \caption{Feynman diagrams generating neutrino transition magnetic moment $\mu_{\nu_e\nu_\mu}$ in the $SU(2)_H$ model. There are additional diagrams where the photon is emitted from the $\tau$ lepton line. The same diagrams with the photon line removed would  contribute to Majorana mass of the neutrino. In the $SU(2)_H$ symmetric limit, the two diagrams add for $\mu_{\nu_e\nu_\mu}$, while they cancel for $m_\nu$.}
    \label{su2h_magnetic}
\end{figure}

The Lagrangian of the model given in Eqs. (\ref{Yuk1})-(\ref{cubic1}) does not respect lepton number.  The $SU(2)_H$ limit of the model however respects $L_e-L_\mu$ symmetry.  This allows a nonzero transition magnetic moment $\mu_{\nu_e \nu_\mu}$, while neutrino mass terms are forbidden -- except for a loop-induced $\tau$ neutrino mass, which will be discussed later.  The Feynman diagrams generating $\mu_{\nu_e\nu_\mu}$ are shown in Fig. \ref{su2h_magnetic}.

 Since the masses of the particles inside the loop are the same in the two diagrams of Fig. \ref{su2h_magnetic}, and since all the couplings are identical, the magnitudes of the two graphs are identical. However, they have a relative minus sign at one of the vertices. When the induced neutrino mass is computed from here, the two diagrams cancel.  On the other hand, with the photon attached to the loop, the two diagrams add (note that the direction of electric charge flow is opposite in the two diagrams).  
 As a result, the two graphs add to give a finite magnetic moment:
\begin{equation}
\mu_{\nu_e\nu_\mu}=\frac{f f^{\prime}}{8 \pi^{2}} m_{\tau} \sin 2 \alpha\left[\frac{1}{m_{h^+}^{2}}\left\{\ln \frac{m_{h^+}^{2}}{m_{\tau}^{2}}-1\right\}-\frac{1}{m_{H^+}^{2}}\left\{\ln \frac{m_{H^+}^{2}}{m_{\tau}^{2}}-1\right\}\right]~.
\label{moment}
\end{equation}
Here $(m_{h^+},\,m_{H^+})$ denote the common masses of the two charged scalars ($h_i^+,\,H_i^+$).

\begin{figure}[h!]
    \centering
      \includegraphics[width=0.6\textwidth]{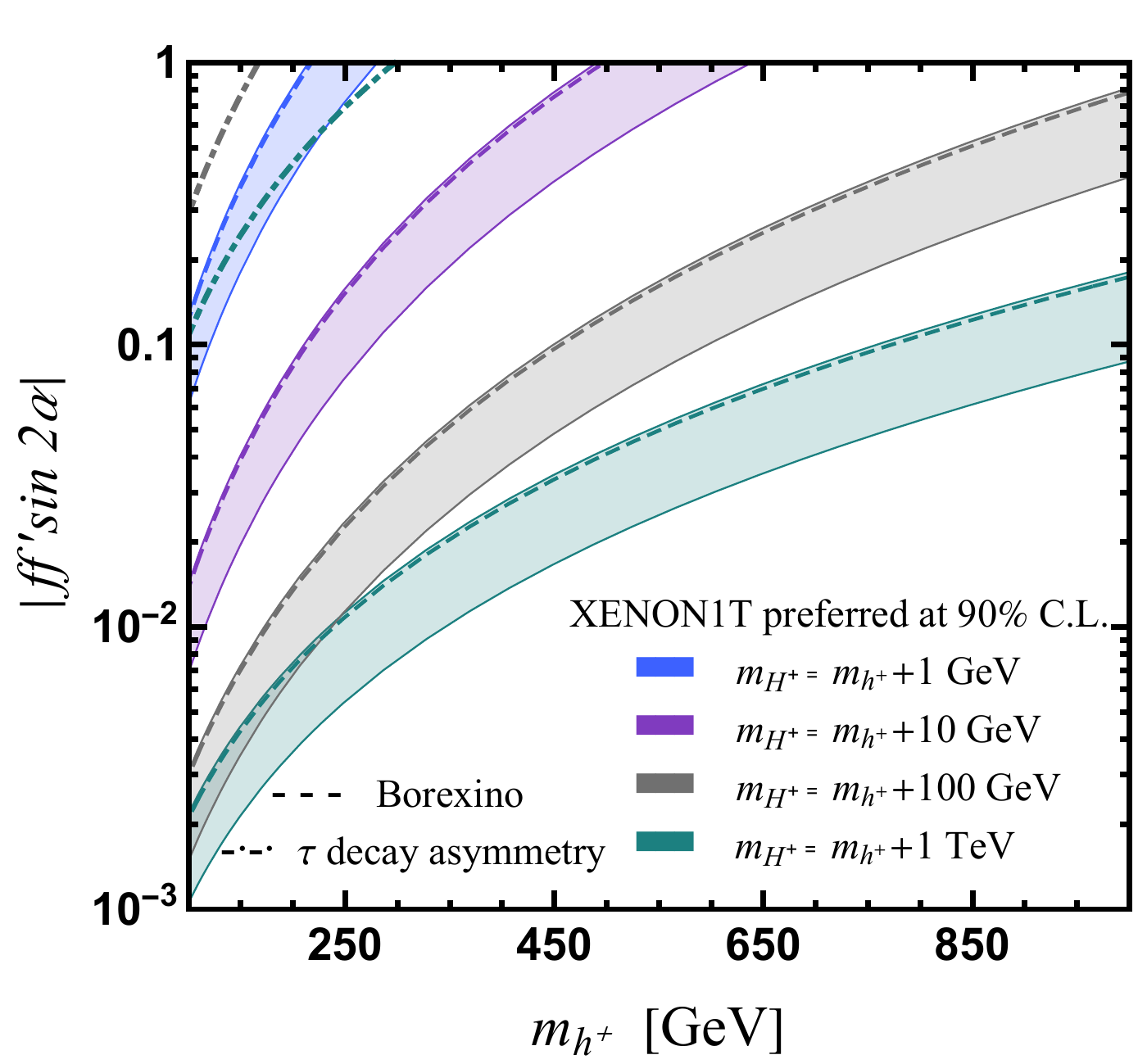}
    \caption{Allowed parameter space in $|ff'\sin{2\alpha}|-m_h^+$ plane to explain XENON1T electron recoil excess.  Blue, purple, gray and green color correspond to different mass choices $m_{H^+} = m_{h^+} + 1$  GeV, $m_{H^+} = m_{h^+} + 10$  GeV, $m_{H^+} = m_{h^+} + 100$  GeV and $m_{H^+} = m_{h^+} + 1$  TeV respectively. The dashed lines indicate the present BOREXINO limit on neutrino magnetic moment and the dot-dashed lines denote the exclusion lines from  $\tau $ decay asymmetry. }
    \label{su2h_sumcon}
\end{figure}

We have plotted contours of constant magnetic moment in the plane of $m_{h^+}$ and $|ff'\sin2\alpha|$ in Fig. \ref{su2h_sumcon}. The different contours represent different values of the heavier charged Higgs mass $m_{H^+}$. Also shown in the figure are the exclusion limit from Borexino on $\mu_{\nu_e\nu_\mu}$ as well as the limit on the parameters from $\tau$ decay asymmetry, discussed later. It can be seen that for $h^+$ masses below about 1 TeV, and for couplings less than one, the model can generate sufficiently large $\mu_{\nu_e\nu_\mu}$ to explain the XENON1T anomaly.  

\begin{figure}[h!]
    \centering
      \includegraphics[width=0.5\textwidth]{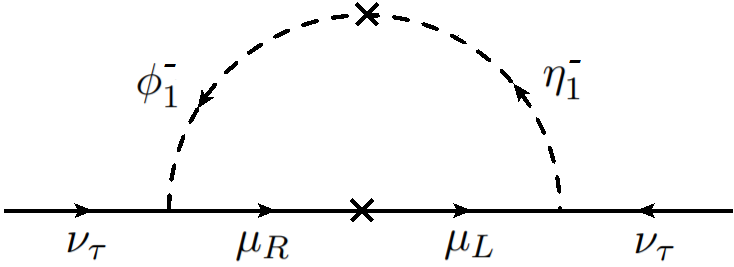}

    \caption{Feynman diagrams generating $\nu_{\tau}$ mass in the $SU(2)_H$ model. }
    \label{su2h_massneu}
\end{figure}

In the $SU(2)_H$ symmetric limit, masses of $\nu_e$ and $\nu_\mu$ are zero.  However, $\nu_\tau$ receives a nonzero mass via the one loop diagram shown in Fig. \ref{su2h_massneu}.  This mass can be evaluated to be
\begin{equation}
    m_{\nu_\tau} = \frac{h_3 f \sin2\alpha}{32\pi^2} m_\mu {\rm ln}\left(\frac{m_{h^+}^2}{m_{H^+}^2} \right)~.
\end{equation}
By choosing $h_3$ to be sufficiently small, this contribution can be made to match neutrino oscillation data, with no adverse effect on the induced $\mu_{\nu_e\nu_\mu}$.  

\subsection{Breaking of \boldmath{$SU(2)_H$} symmetry and induced neutrino mass}

$SU(2)_H$ cannot be exact, as it would imply $m_e=m_\mu$.  We propose to include explicit but small breaking of $SU(2)_H$, so that realistic electron and muon masses can be generated.\footnote{If $m_\mu \neq m_e$ is realized by the VEVs of $\langle \phi_1^0\rangle \neq \langle \phi_2^0\rangle$, the mixing angles amongst $\mu_L-\tau_L$, as well as $e_R-\tau_R$ will become relatively large, leading to unacceptably large $m_\nu$ when $\mu_\nu$ is demanded to be large.}  Since the breaking we introduce is hard, via dimension four terms in the Lagrangian, to be consistent, other renormalizable $SU(2)_H$ breaking terms should also be included.  Neutrino masses will then be induced proportional to these $SU(2)_H$ breaking terms. Here we  show that such breaking terms of the order needed to achieve $e-\mu$ mass splitting induce $m_\nu$ of the right order of magnitude to explain neutrino oscillation data.

The Yukawa couplings of Eq. (\ref{yukawasu2}) will receive explicit $SU(2)_H$ breaking corrections given by
\begin{eqnarray}
\Delta \mathcal{L}_{\rm Yuk} &=& \delta h_{1}\left[(\bar{\nu}_{e} e_{R} \phi_{S}^{+}+\bar{e}_{L} e_{R} \phi_{S}^{0})-(\bar{\nu}_{\mu} \mu_{R} \phi_{S}^{+}+\bar{\mu}_{L} \mu_{R} \phi_{S}^0)\right] 
 \nonumber\\
&+& \delta h_{3}\left[-(\bar{\nu}_{\tau} e_{R} \phi_{2}^{+} +\bar{\tau}_{L} e_{R} \phi_{2}^{0}) - (\bar{\nu}_{\tau} \mu_{R} \phi_{1}^{+}+\bar{\tau}_{L} \mu_{R} \phi_{1}^{0})\right]\nonumber\\
&+& \delta f\left[\left(\nu_{e}^T C \tau_{L} - e_{L}^{T} C \nu_\tau\right) \eta_{2}^{+} + \left(\nu_{\mu}^{T} C \tau_{L}-\mu_{L}^{T} C \nu_{\tau} \right) \eta_{1}^{+}\right]\nonumber\\
&+&\delta f^{\prime}\left[(\bar{\nu}_{e} \tau_{R} \phi_{1}^{+}+\bar{e}_{L} \tau_{R} \phi_{1}^{0}) - (\bar{\nu}_{\mu}  \tau_{R} \phi_2^{+}+\bar{\mu}_{L} \tau_{R} {\phi_2^{0}})\right] + H.c. \label{yukawasu2prime}
\end{eqnarray}
The $\delta h_1$ term would split the masses of the electron and the muon, since their masses are now given by
\begin{equation}
    m_e = (h_1 + \delta h_1) \,\frac{v}{\sqrt{2}} \equiv h_e \,\frac{v}{\sqrt{2}},~~~ m_\mu = (h_1 - \delta h_1)\, \frac{v}{\sqrt{2}} \equiv h_\mu \, \frac{v}{\sqrt{2}}.
\end{equation}
We therefore can express $h_1$ and $\Delta h_1$ in terms of the Standard Model Yukawa couplings $h_e$ and $h_\mu$ as
\begin{equation}
    h_1 = \frac{h_e + h_\mu}{2},~~~ \delta h_1 = \frac{h_e-h_\mu}{2}
\end{equation}
which can be used to explicitly show $SU(2)_H$ breaking. Note that Eq. (\ref{yukawasu2prime}), along with Eq. (\ref{yukawasu2}) forms the most general set of Yukawa couplings consistent of the theory. (We have not written down $L_e-L_\mu$ violating terms in the Lagrangian, since the masses of the electron and muon do not break this symmetry.)

Similarly, the scalar potential of Eq. (\ref{potential}) should include the following terms that break $SU(2)_H$ explicitly:
\begin{eqnarray}
\Delta V &\supset& \delta m_\eta^2(|\eta_1|^2-|\eta_2|^2) +\delta  m_{\phi^+}^2 ( |\phi_1^+|^2 - |\phi_2^+|^2) + \delta  m_{\phi^0}^2 (|\phi_1^0|^2 - |\phi_2^0|^2) \nonumber \\
&+& \delta \mu\left[\overline{\phi_S^0} \left(\eta_{1}^{+} \phi_{1}^{-}-\eta_{2}^{+} \phi_{2}^{-}\right)-\phi_{S}^{-}\left(\eta_{1}^{+} \bar{\phi}_{1}^{0}-\eta_{2}^{+} \bar{\phi}_{2}^{0}\right)\right] + H.c.
\label{potentialprime}
\end{eqnarray}
Together with Eq. (\ref{potential}), this becomes the most general potential relevant for neutrino mass/magnetic moment discussions.

\begin{figure}[h!]
    \centering
      \includegraphics[width=0.6\textwidth]{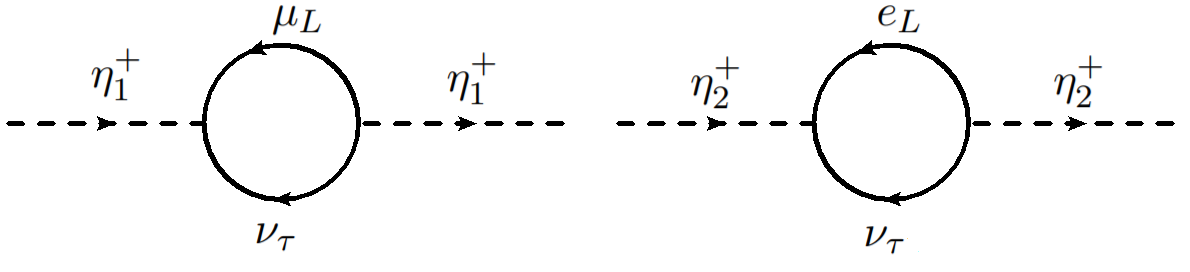}
    \caption{Feynman diagrams generating mass splitting among doublet scalars of  $SU(2)_H$. }
    \label{su2h_breaking1}
\end{figure}

We now proceed to compute the shifts in the couplings and masses induced by the electron and muon masses and Yukawa couplings.  The diagrams shown in Fig. \ref{su2h_breaking1} would lead to a splitting in the masses of $\eta_1^+$ and $\eta_2^+$, once the Yukawa interactions of Eq. (\ref{yukawasu2prime}) are included. We evaluate these diagrams in dimensional regularization, adopt a zero momentum subtraction scheme to determine the mass and wave function counter-terms, and obtain the finite shift in the mass of $\eta_1^+$ as
\begin{equation}
\delta m_{\eta_1}^2 = \frac{|f|^2}{8\pi^2} m_\eta^2 \left( -\frac{1}{4} + \frac{1}{2} \, {\rm ln}\left(\frac{m_\eta^2}{m_\mu^2} \right) \right)~.
\end{equation}
A similar expression is obtained for the renormalized mass correction of $\eta_2^+$ with $m_\mu$ replaced by $m_e$, resulting in a shift in the two masses given by
\begin{equation}
\delta m_{\eta}^2 = m_{\eta_1}^2 - m_{\eta_2}^2 = \frac{|f|^2}{16 \pi^2}\, m_{\eta}^2  {\rm ln} \left(\frac{m_e^2}{m_\mu^2}  \right)~.    
\end{equation}
A similar calculation for the mass splitting of $\phi_1^+$ and $\phi_2^+$ shows
\begin{equation}
\delta m_{\phi}^2 = m_{\phi_1}^2 - m_{\phi_2}^2 = \frac{|h_3|^2}{16 \pi^2} \,  m_{\phi}^2 {\rm ln} \left(\frac{m_e^2}{m_\mu^2}  \right)~.    
\end{equation}

Since the dependence on the leptons masses in these mass splittings is only logarithmic, in order to achieve $\delta m^2/m^2 \sim 10^{-5}$ needed to ensure the smallness of neutrino masses, it is necessary to take $|f| \sim {\rm few} \times 10^{-3}$, which is consistent with the needed $\mu_{\nu_e\nu_\mu}$, see Fig. \ref{su2h_sumcon}. 

\begin{figure}[h!]
    \centering
      \includegraphics[width=0.6\textwidth]{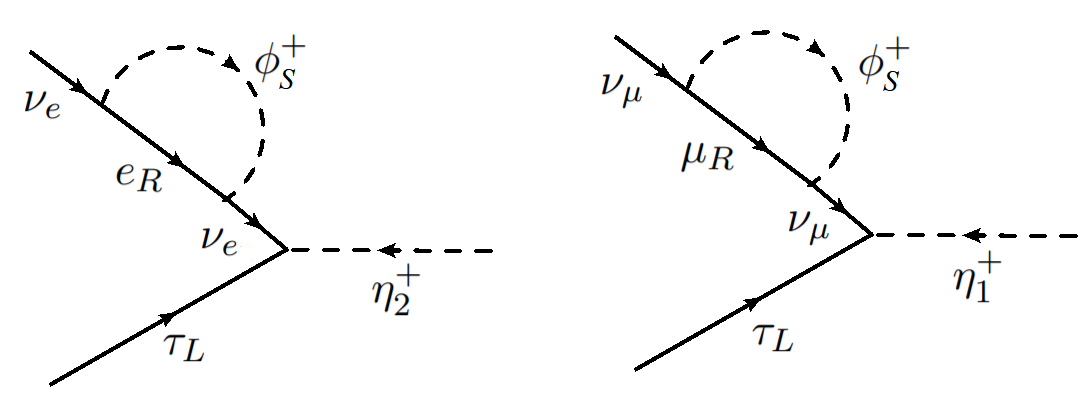}
      \includegraphics[width=0.6\textwidth]{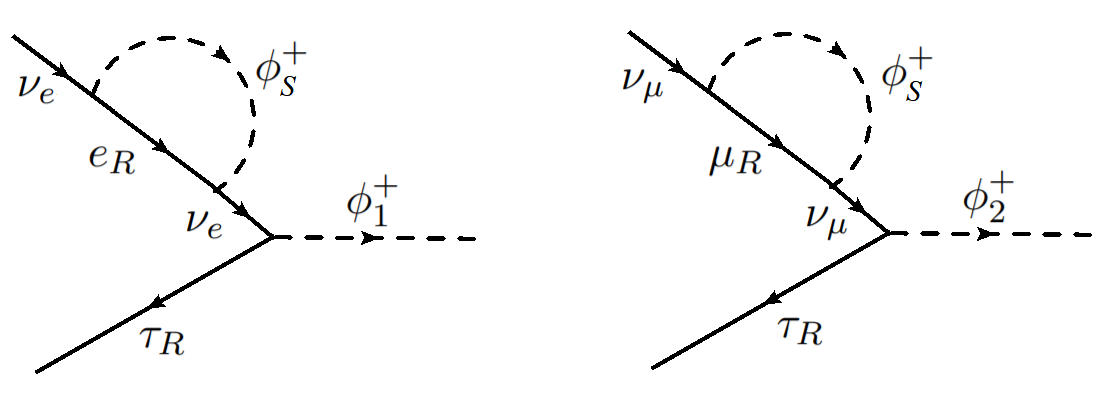}
    \caption{Feynman diagrams generating coupling shifts in the $SU(2)_H$ model. }
    \label{su2h_breaking2}
\end{figure}

The first set of Feynman diagrams of Fig. \ref{su2h_breaking2} would result in a shift in the coupling $\delta f$, which is however proportional to the electron and the muon Yukawa couplings:
\begin{equation}
\delta f \simeq \frac{f(h_e^2-h_\mu^2)}{16 \pi^2}
\end{equation}
and sufficiently small to keep the neutrino mass within the observed range.  
Similarly, a shift in the $f'$ couplings is induced by  the second set of diagrams in Fig. \ref{su2h_breaking2}, given by
\begin{equation}
\delta f' \simeq \frac{f'(h_e^2-h_\mu^2)}{16 \pi^2}~.
\end{equation}

\begin{figure}[h!]
    \centering
    \includegraphics[width=0.6\textwidth]{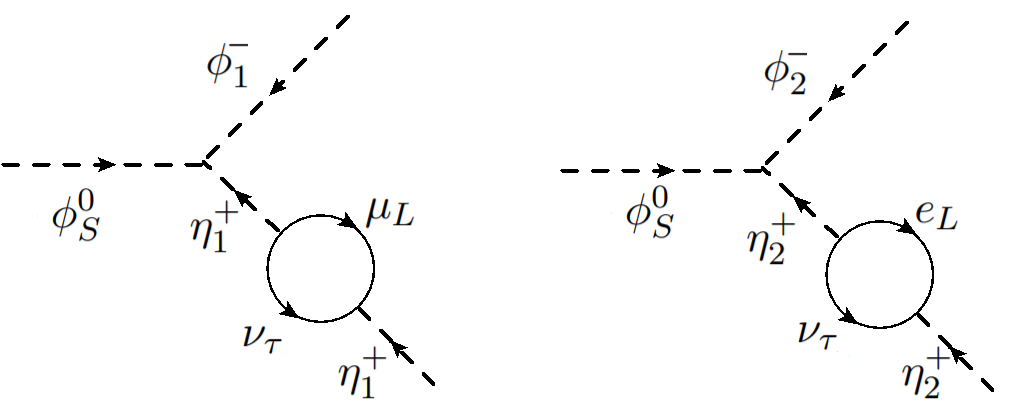}
    \caption{Feynman diagrams generating shift in the cubic scalar couplings in the $SU(2)_H$ model. }
    \label{su2h_breaking3}
\end{figure}

The Feynman diagrams of Fig. \ref{su2h_breaking3} will induce shifts in the cubic scalar couplings, which we estimate to be
\begin{equation}
\delta \mu \simeq  \frac{|f|^2 \mu}{16 \pi^2}\, {\rm ln} \left(\frac{m_e^2}{m_\mu^2}  \right)~.    
\end{equation}
This shift is also not excessive, for $|f| \simeq 10^{-2}$.  

Once the masses of ($\eta_1^+,\, \eta_2^+)$ split, and similarly with $(\phi_1^+,\,\phi_2^+)$, nonzero neutrino masses will be induced.  Keeping the largest splittings into account the induced $m_{\nu_e\nu_\mu}$ is given by
\begin{equation}
    m_{\nu_e \nu_\mu} = \frac{f f' m_\tau \sin 2\alpha}{16 \pi^2}\left[ \frac{\delta m_{H^+}^2}{m_{H^+}^2}  - \frac{\delta m_{h^+}^2}{m_{h^+}^2} + 2 (\delta \alpha)\cot2 \alpha\, {\rm ln} \frac{m_{h^+}^2}{m_{H^+}^2} \right]~.
\end{equation}
Here $\delta \alpha$ is the shift in the mixing angle arising primarily from the shift in the $\mu$ term.  For $ f \sim 10^{-2}, \, f' \sim 1, \, \delta m^2/m^2 \sim 10^{-5}$, one obtains $m_{\nu_e \nu_\mu} 
\sim 0. 1$ eV, which is roughly the value needed to explain neutrino oscillation data. As for other entries in the neutrino mass matrix, one could introduce $SU(2)_H$ violating small couplings in the $f$ and $f'$ matrices, so that the full mass matrix explains the entirety of neutrino oscillation data.

\section{Phenomenological implications of \boldmath{$SU(2)_H$} model }\label{SEC-05}

In this section, we discuss various phenomenological implications of the $SU(2)_H$ model that induces large $\mu_\mu$ with a suppressed $m_\nu$.  One important consequence is modification of the $\tau$ decay, which provides constraints on the model parameter.  It should be noted that in spite of having light scalars in the theory, there are no significant flavor violating processes in the charged lepton sector, owing to the approximate $SU(2)_H$ symmetry. We derive constraints from LEP experiment, and discuss prospects of testing the model at high luminosity LHC.

%%%%%%%%%%%%%%%%%%%%%%%%%%%%%%%%%%%%%%%%%
\subsection{Modifications to \boldmath{$\tau$} decay} 
%%%%%%%%%%%%%%%%%%%%%%%%%%%%%%%%%%%%%%%%%

The new charged scalars present in the $SU(2)_H$ model would contribute to the normal leptonic decays of the $\tau$ lepton.  They would also lead to new lepton number violating decays.  There are constraints on the model parameters arising from $\tau$ lifetime as well as asymmetry parameters in the decay. The relevant Yukawa Lagrangian is given in Eq. (\ref{yukawasu2}). It is sufficient to work in the $SU(2)_H$ symmetric limit, since violations of $SU(2)_H$ will be very small.

\begin{figure}[h!]
    \centering
      \includegraphics[width=0.35\textwidth]{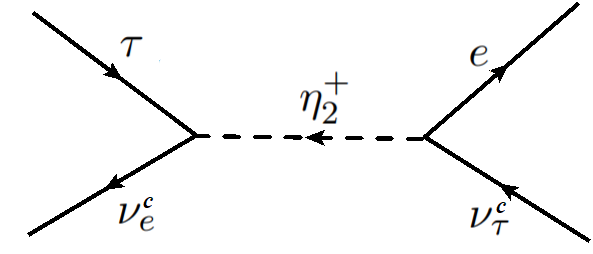}\hspace{0.5in}
        \includegraphics[width=0.35\textwidth]{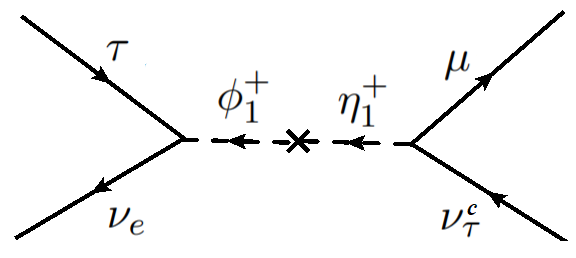}
    \caption{Feynman diagrams contributing  to $\tau$ decay via the exchange of charged scalars in the $SU(2)_H$ model. The diagram on the left contributes to the normal $\tau$ decay, $\tau \rightarrow e \nu_\tau \overline{\nu}_e$.  There is a similar diagram mediated by $\eta_2^+$ leading to $\tau \rightarrow \mu \nu_\tau \overline{\nu}_\mu$ with identical strength. The diagram on the right contributes to a new decay channel, $\tau \rightarrow \mu \nu_e \nu_\tau$, with an analogous diagram mediated by $\phi_2^+=\eta_2^+$ for $\tau \rightarrow e \nu_\mu \nu_\tau$. }
    \label{taudecay}
\end{figure}

The charged scalar $\eta_2^+$ mediates the decay  $\tau \rightarrow e \nu_\tau \overline{\nu}_e$, as shown in Fig. \ref{taudecay}, left panel.  There is an identical diagram for the decay  $\tau \rightarrow \mu \nu_\tau \overline{\nu}_\mu$ mediated by $\eta_1^+$.  These two amplitudes being the same, there is no lepton universality violation in $\tau$ decays.  The leptonic decay rates will be modified as
\begin{equation}
    \Gamma_{\tau \rightarrow \ell \overline{\nu}_\ell \nu_\tau} = \Gamma_{SM} (1+ \epsilon_\tau)^2
\end{equation}
where
\begin{equation}
    \epsilon_\tau = \frac{f^2}{g^2}m_W^2\left(\frac{\cos^2\phi}{m_{h^+}^2} + \frac{\sin^2\phi}{m_{H^+}^2} \right)~.
\end{equation}
The $\tau$ lifetime has been measured accurately to be $\tau_{\rm exp} = (290.75 \pm 0.36) \times 10^{-15}\, {\rm sec.}$, while the Standard Model prediction for the lifetime is $\tau_{SM} = (290.39 \pm 2.17) \times 10^{-15}\, {\rm sec.}$ Adding the two errors in quadrature we find new contributions should be limited to 1.64\%. Noting that the new contributions only affect the leptonic modes that has a branching ratio of 17.8\%, we obtain $\epsilon_\tau < 0.0227$.  This constraint is rather easy to satisfy within the model, and has been imposed in our calculation of $\mu_{\nu_e\nu_\mu}$.

The right panel of Fig, \ref{taudecay} shows new lepton number violating decays of the $\tau$, $\tau \rightarrow \mu \nu_e \nu_\tau$.  There is a similar diagram for the decay $\tau \rightarrow e \nu_\mu \nu_\tau$ mediated by $\eta_2^+-\phi_2^+$ exchange. Note that these processes conserve $L_e-L_\mu$, which is an apprixmate symmetry of the model.  Thee new decays do not interfere with the standard decay, but can modify the lifetime and decay asymmetry parameters.  The effective Lagrangian for the decay is found to be
\begin{equation}
    {\cal L}_{\rm eff} = -\frac{1}{2} f f' \sin2\phi \left(\frac{1}{m_{h^+}^2}-\frac{1}{m_{H^+}^2}  \right) (\overline{\mu}_L \nu_e^c) (\overline{\nu}_e \tau_R)~.
\end{equation}
Defining $\epsilon_\tau'$ as
\begin{equation}
    \epsilon_\tau' = \frac{f f'}{2 g^2}\sin2\phi\left(\frac{1}{m_{h^+}^2}-\frac{1}{m_{H^+}^2} \right),
\end{equation}
we see that $\Gamma_\tau^{\rm new} = \Gamma_\tau^{SM}(1 + {\epsilon'}_\tau^2)$.  This would lead to a limit of $|\epsilon_\tau'| < 0.205$ from $\tau$ lifetime.

Comparing with muon decay formalism of Ref. \cite{Kuno:1999jp}, we see that for the lepton number violating $\tau$ decays, $g_{LR}^S = 2 \epsilon_\tau'$.  This leads to the following modifications of the asymmetry parameter in $\tau$ decay \cite{Babu:2016fdt}:
\begin{eqnarray}
        \rho = \frac{3}{4}, ~~~\delta = \frac{3}{4}, \xi = 1 - 2|\epsilon_{\tau}'|^2,~~~\delta \,\xi = \frac{3}{4}(1-2|\epsilon'_\tau|^2)~.
\end{eqnarray}
Using the experimental value $\delta \, \xi = 0.746 \pm 0.021$ \cite{Tanabashi:2018oca}, and using 2 $\sigma$ error bar, we obtain  $|\epsilon_\tau'| < 0.175$.  The constraint from the measurement $\xi = 0.985 \pm 0.030$, $|\epsilon_\tau'|< 0.194$ is somewhat weaker.

The parameters that enter the new $\tau$ decay are the same as for the neutrino magnetic moment.  We have indicated the most stringent constraint, arising from $\delta\,\xi$ in Fig. \ref{su2h_sumcon}.  While this does provide a useful constraint, large magnetic moment of the neutrino is still permitted by this constraint.

%%%%%%%%%%%%%%%%%%%%%%%%%%%%%%%%%%%%%%%%%

%%%%%%%%%%%%%%%%%%%%%%%%%%%%%%%%%%%%%%%%%
\subsection{LEP constraints} 
%%%%%%%%%%%%%%%%%%%%%%%%%%%%%%%%%%%%%%%%%
\begin{figure}[h!]
    \centering
      \includegraphics[width=0.22\textwidth]{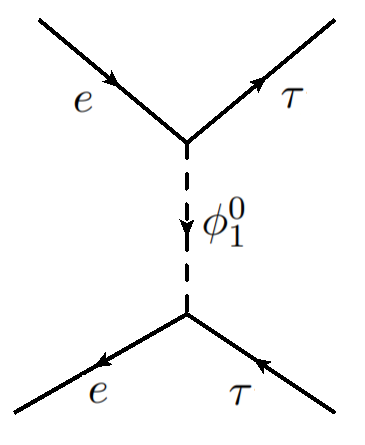}
    \caption{Feynman diagram contributing to the process $e^- e^+ \to \tau^- \tau^+$ process in the $SU(2)_H$ model.}
    \label{lep1}
\end{figure}

At LEP, the $t$-channel exchange of the neutral component of the $\Phi$ multiplet $\left(\phi_{1}^{0} \right)$  can contribute significantly to the process $e^{+} e^{-} \rightarrow \tau^{+} \tau^{-}$ as shown in Fig.~\ref{lep1}. Contact interactions involving $e^+ e^-$ and a pair of fermions~\cite{LEP:2003aa} are tightly constrained by the LEP experiments. For heavy mass of $\phi_{1}^{0}$, one can integrate it out and express its effect via a dimension-6 operator\footnote{Contact interactions are parametrized as an effective Lagrangian for the process $e^+ e^- \to \tau^+ \tau^-$: \cite{Eichten:1983hw}
\begin{equation}
{\cal L}_{\rm eff}\  = \ \frac{g^2}{{\Lambda}^2} \sum_{i,j=L,R} {\eta}_{ij}^{\tau}  (\bar e_i \gamma^\mu  e_i)(\bar \tau_j \gamma_\mu  \tau_j)\, ,  \label{eq:eff}
\end{equation}
where  $\Lambda$ is the new physics scale  and $\eta^\tau_{ij}=\pm 1$ or 0, depending on the chirality structure.}.  Therefore, the LEP constraint on the scale $\Lambda$ of the contact interaction  for the process $e^{+} e^{-} \rightarrow \tau^{+} \tau^{-}$, viz., $\Lambda > 2.2$ TeV, can be translated to a limit as  $m_{\phi_{1}^0}/|f'|>0.44$ TeV. However, if  the $\phi_{1}^{0}$ scalar is lighter than about 300 GeV, the LEP contact interaction limit is not applicable. For lighter $\phi_{1}^{0}$ scalar
we compute the cross-section at the parton-level for the process  $e^+e^-\to \tau^+\tau^-$  using {\tt MadGraph5aMC@NLO} event generator~\cite{Alwall:2014hca} and  compare it with the measured cross sections~\cite{LEP:2003aa, Abbiendi:2003dh}, imposing identical acceptance criteria \cite{Abbiendi:2003dh} and obtain limits on the Yukawa coupling $f'$ as a function of the mass $m_{\phi_{1}^0}$. At $\sqrt{s}=207$ GeV, with an integrated luminosity of 134.5 pb$^{-1}$, LEP observed a total of 206 events \cite{Abbiendi:2003dh} for the process $e^{+} e^{-} \rightarrow \tau^{+} \tau^{-}$, which can be translated into a limit on the cross-section of $7.21 \pm 0.57 \pm 0.19$ pb (the first error shown is statistical and the second one is systematic). The SM predicted cross-section is 7.12 pb. Comparing this cross-section bound at $2\sigma$ level, we find that for the benchmark  values of the $\phi_1^0$ mass  $m_{\phi_{1}^0} = 100$ and 200 GeV, the Yukawa coupling $f'$ can be as big as 0.675 and 0.925  respectively. It should be mentioned that the other new neutral scalar field $\phi_{2}^{0}$ has no impact on the LEP experiment as, it couples to $\mu $ and $\tau$ leptons.

There are four physical charged scalars ($\eta_1^+,\eta_2^+, \phi_1^+, \phi_2^+$) in the $SU(2)_H$ model. (Here we neglect for simplicity of discussion the mixings of $\eta_1^+-\phi_1^+$ as well as $\eta_2^+-\phi_2^+$.) At LEP, these charged scalars  can be pair-produced  via Drell-Yan process with the exchange of $\gamma$ or  $Z$ boson in the $s$ channel. $\eta_2^+$ can also be  pair-produced through the $t$-channel $\nu_{\tau}$ exchange. Due to the absence of  flavor-diagonal Yukawa couplings of the charged scalars, they cannot be produced in association with $W$ boson, which relaxes various LEP search limits such as $c\bar{s}\tau \nu$ searches. Once produced on-shell, the charged scalar and the neutral scalar would decay into  various leptonic final states with the dominant decay modes given by
\begin{equation}
\begin{array}{l}
\eta_1^+ \rightarrow  \mu \nu_{\tau}, \tau \nu_{\mu}, \quad
\eta_2^+ \rightarrow  e \nu_{\tau}, \tau \nu_e, \quad \phi_1^+ \rightarrow  \tau \nu_e, \quad
\phi_2^+ \rightarrow \tau \nu_{\mu}, \quad
\phi_1^0 \rightarrow  e  \tau, \quad
\phi_2^0 \rightarrow \mu \tau.
\end{array}
\end{equation}
For our numerical analysis, we compute  the cross-sections using {\tt MadGraph5aMC@NLO} event generator~\cite{Alwall:2014hca} at the parton level. There are several supersymmetirc slepton searches ~\cite{lepsusy} at LEP, which we reinterpret  ~\cite{lepsusy} as limits on charged Higgs particles since they  mimic these signatures. In our model, the decay branching ratios of $\eta_1^+$  to  $\mu \nu_{\tau}$ and $\tau \nu_{\mu}$ modes are equal and $50\%$. Hence, $\eta_1^+$ needs to obey both the smuon and stau search limits. The smuon searches impose a limit on  $\eta_1^+$ wich allowed to be as low as 91 GeV. Similarly, $\eta_2^+$ decays to  $e \nu_{\tau}$ and $\tau \nu_{e}$ modes equally. Selectron and stau search impose limit on the mass of $\eta_2^+$, and we find the more severe limit of $m_{\eta_2^+} > 96$ GeV  arising from the selectron searches.   We also find lower limits on the masses of $\phi_1^+$ and $\phi_2^+$ to be 88 GeV, originating mainly from stau searches since both decay to $\tau \nu_{e}$ and $\tau \nu_{\mu}$ 100$\%$ times.  We conclude that in $SU(2)_H$ model, a light charged scalar with mass as low as $\sim$88 GeV is consistent with all LEP search limits.

%%%%%%%%%%%%%%%%%%%%%%%%%%%%%%%%%%%%%%%%%
\subsection{LHC prospects} 
%%%%%%%%%%%%%%%%%%%%%%%%%%%%%%%%%%%%%%%%%
\begin{figure}[h!]
    \centering
      \includegraphics[width=0.6\textwidth]{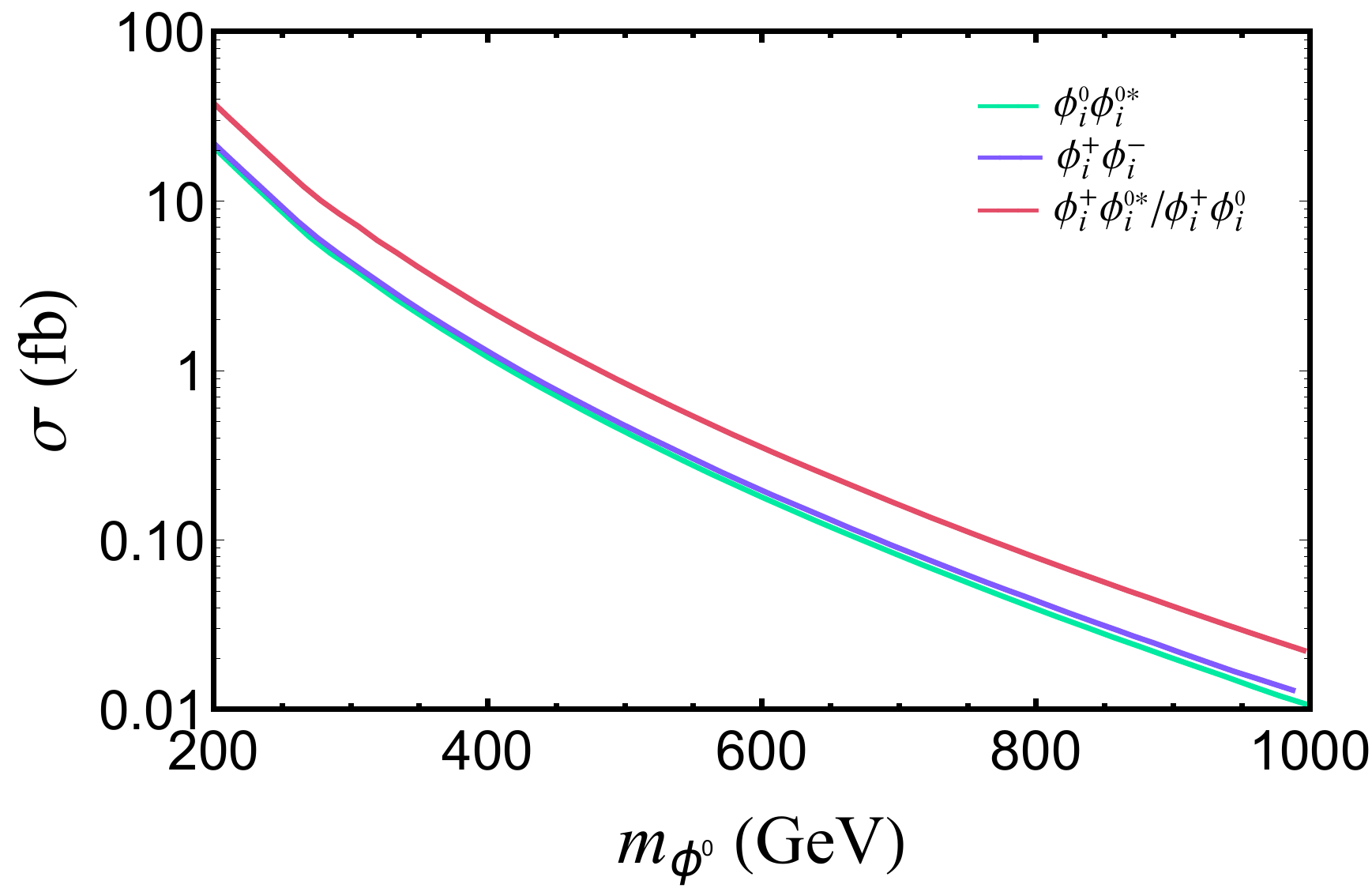}
    \caption{Di-Higgs production cross-section at the LHC in the $SU(2)_H$ model.}
    \label{lhc1}
\end{figure}

In our $SU(2)_H$ model, the neutral scalars $\phi_1^0$ and $ \phi_2^0$ do not couple to quarks, and hence, cannot be directly produced via the gluon fusion process.  The $CP-$ even neutral scalars Re($\phi_1^0$) and Re($\phi_2^0$)  will be dominantly produced in association with their $CP-$ odd partners Im($\phi_1^{0}$) and Im($\phi_2^{0}$) via $s-$ channel $Z$ boson exchange   in quark fusion processes at the LHC. The dominant production mechanism for the charged scalars ($\phi_i^+$) at the LHC will be via $s-$ channel $Z/\gamma $ exchange. In addition, the neutral scalars $\phi_1^0$ and $ \phi_2^0$ can  be produced in association with their charged partners $\phi_1^+$ and $ \phi_2^+$  via $s-$ channel $W$ boson exchange. The production cross-sections at the 14 TeV LHC  for the processes $pp \to \phi_i^0 \phi_i^{0*},\phi_i^0 \phi_i^{+}, \phi_i^+ \phi_i^{0*}, \phi_i^+ \phi_i^{-} $ are shown in Fig.~\ref{lhc1}. Searches for a heavy neutral at the LHC in the context of either 2HDM (two Higgs doublet model) or MSSM (minimal supersymmetric standard mode) are not directly applicable to our scenario since $\phi_1^0$ and $ \phi_2^0$ do not couple to quarks.
There has been searches for sleptons produced in pairs at $\sqrt s=13$ TeV LHC in the mass range above a 100 GeV.  We found that the current limits \cite{Sirunyan:2018vig}  on these cross section are larger than the $\phi_i^+\phi_i^-$ Drell-Yan pair-production rate, and hence there are no stringent  limits for these leptophilic charged scalars from the LHC. 

 \begin{figure}[ht!]
    \centering
      \includegraphics[width=0.5\textwidth]{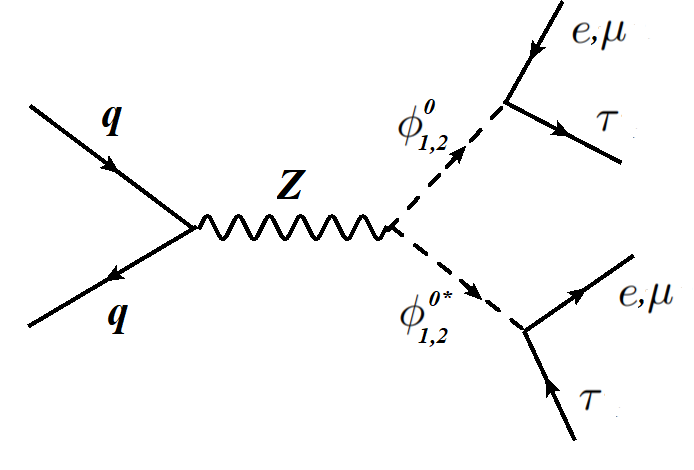}
    \caption{Feynman diagram for the novel signature $p p \to e^- e^+ \tau^- \tau^+, \,\mu^- \mu^+ \tau^- \tau^+$ in the $SU(2)_H$ model at the LHC via exchange of the $CP$ even and $CP$ odd scalars.}
\label{col1}
\end{figure}

\begin{table}[t!]
\centering
\resizebox{0.95\textwidth}{!}{%
\begin{tabular}{||c|c|c|c|c|c|c|c||}
\hline
\hline
\multicolumn{8}{||c||}{\bf Number of events of the signal and SM background  at 14 TeV LHC }                                                                                                   \\ \hline
\multirow{10}{*}{\textbf{SM Background}} &
  Dataset &
  $N_{l} \geq 4$ $\&$ $N_{\tau_{h}} \geq 2$ &
  $N_{b}=0$ &
  $\mathrm{TTHE} \leq 20 $ GeV&
  $W$-veto &
  $Z$ -veto &
  $\frac{\mathrm{S}}{\sqrt{S+B}}$ \\ \cline{2-8} 
 & $V V_{l}$       & $90$   & $90$   & $45$     & $45$       & $45$       & $-$           \\ \cline{2-8} 
 & $ t V_{l}$         & $225$      & $210$      & $120$       & $120$       & $120$       & $-$           \\ \cline{2-8} 
 & $ t \bar{t} V_{l}$ & $18$  & $9$      & $1.8$       & $1.8$       & $1.8$       & $-$           \\ \cline{2-8} 
 & $t V V_{l}$      & $15$  & $9$  & $1.5$ & $1.5$ & $1.5$ & $-$           \\ \cline{2-8} 
 & $\bar{t} t_l $             & $4590$      & $2610$      & $210$       & $0$       & $0$       & $-$           \\ \cline{2-8} 
 & $t \bar{t} h$               & $45$      & $0$  & $0$       & $0$       & $0$       & $-$           \\ \cline{2-8} 
 & $t \bar{t} t \bar{t}$             & $12$     & $0$     & $0$       & $0$       & $0$       & $-$           \\ \cline{2-8} 
 & $V V V$           & $30$  & $12$  & $0$ & $0$ & $0$ & $-$           \\ \cline{2-8} 
 & Total                 & $5025$   & $2940$   & $378$     & $168$      & $168$      & $-$           \\ \hline \hline
\multirow{5}{*}{\begin{tabular}[c]{@{}c@{}}\textbf{Signal}\\ $SU(2)_H$ Model\end{tabular}} &
   $m_{\phi^0}$ = 300 GeV &
  $1197$ &
  $1177$ &
  $585$ &
  $499$ &
  $492$ &
  $19.1$ \\ \cline{2-8} 
 & $m_{\phi^0}$ = 400 GeV        & $350$     & $344$     & $171$      & $146$      & $144$      & $8.1$ \\ \cline{2-8} 
 & $m_{\phi^0}$ = 500 GeV        & $138.6$     & $136.2$     & $67.8$      & $58$      & $57$      & $3.8$ \\ \cline{2-8} 
 & $m_{\phi^0}$ = 600 GeV        & $53.55$     & $52.63$     & $26.16$      & $22.34$      & $22.05$      & $1.6$ \\ \cline{2-8} 
 &  $m_{\phi^0}$ = 800 GeV        & $12.6$ & $12.39$ & $6$   & $5.25$   & $5.19$   & $0.4$ \\ \hline \hline
\end{tabular}%
}
\caption{Collider analysis of the signal $pp \to l^+ l^- \tau^+ \tau^-$ in $SU(2)_H$ model. The expected number of signal events and SM background are presented, together with the effect of a set of acceptance cuts, for LHC center-of-mass energy $\sqrt{s}=$ 14 TeV  with integrated luminosity  $\mathcal{L}=300$ fb$^{-1}$. The index $V$ denotes the SM gauge bosons and  subscript $l$ in the dataset indicates leptonic decay of gauge bosons and top quarks. Signal events for five different masses $m_{\phi_i^0}$= 300, 400, 500, 600 and 800 GeV are analyzed.}
\label{tab:lhcana}
\end{table}

The most promising signal of the model is $p p \to e^- e^+ \tau^- \tau^+,\, \mu^- \mu^+ \tau^- \tau^+$ at the LHC as shown in Fig.~\ref{col1}. Once produced, $\phi_i^0$ would decay into a pair of different flavored leptons; i.e., $\phi_1^0$ decays to $e \tau$ and $\phi_2^0$ decays to $\mu \tau$. Since final state leptons with large transverse momenta can be  identified cleanly with good resolution at the LHC, this signal will be a good test for this model. Although there are several experimental multi-lepton searches\cite{Sirunyan:2018qlb, Aaboud:2018puo},  most of them assume a heavy $ZZ^{(\star)}$ resonance~\cite{Sirunyan:2018qlb, Aaboud:2018puo}, which are not applicable to our scenario. In the context of supersymmetric models, there are inclusive multilepton searches, mostly with large missing transverse energy~\cite{Chatrchyan:2014aea, Aaboud:2018zeb}. There are no dedicated searches for the $l^+l^-\tau^+\tau^-$ process. However, in the context of sneutrino searches, this type of signal could arise for specific scenarios explored in Ref. \cite{Ghosh:2017yeh}. The SM background for the $pp \to l^+  l^- \tau^+ \tau^-$ process is principally from  the pair production of gauge bosons $ZZ, \,WW$ and $ZW$; from the top quark production through the channels $\bar{t} t$ and $\bar{t} t \bar{t} t$;  the associated production  $\bar{t} t V, t V$ and $t V V$; and the Higgs production in association with top quark pair. We have analyzed all these processes and summarized the results in Table~\ref{tab:lhcana}. We note  several distinguishing characteristics of the signal of the $SU(2)_H$ model: (a) the invariant mass distributions for different flavor lepton pair from the $\phi_i^0$ decay would peak at a value different from the $Z$ boson mass; (b)  if it is originated from Higgs signal, $h \to ZZ^*$ should be accompanied by $h \to WW^*$, with a ratio of about 1 to 2; (c) $Z$ decay also leads to neutrino decay modes that are absent in our scenario; (d)
since the process does not involve quarks,  the signal events suffer from significantly smaller  hadronic activity than the associated background events including a leptonically decaying top $t_{l}$; (e)  since the leptons are produced from heavy particle $\phi_i^0$, they are expected to be more energetic than the ones produced in the decay of gauge bosons. We analyze the signal and show that there is a huge potential  to unravel this  multilepton signal above the SM background. 

 \begin{figure}[h!]
    \centering
      \includegraphics[width=0.65\textwidth]{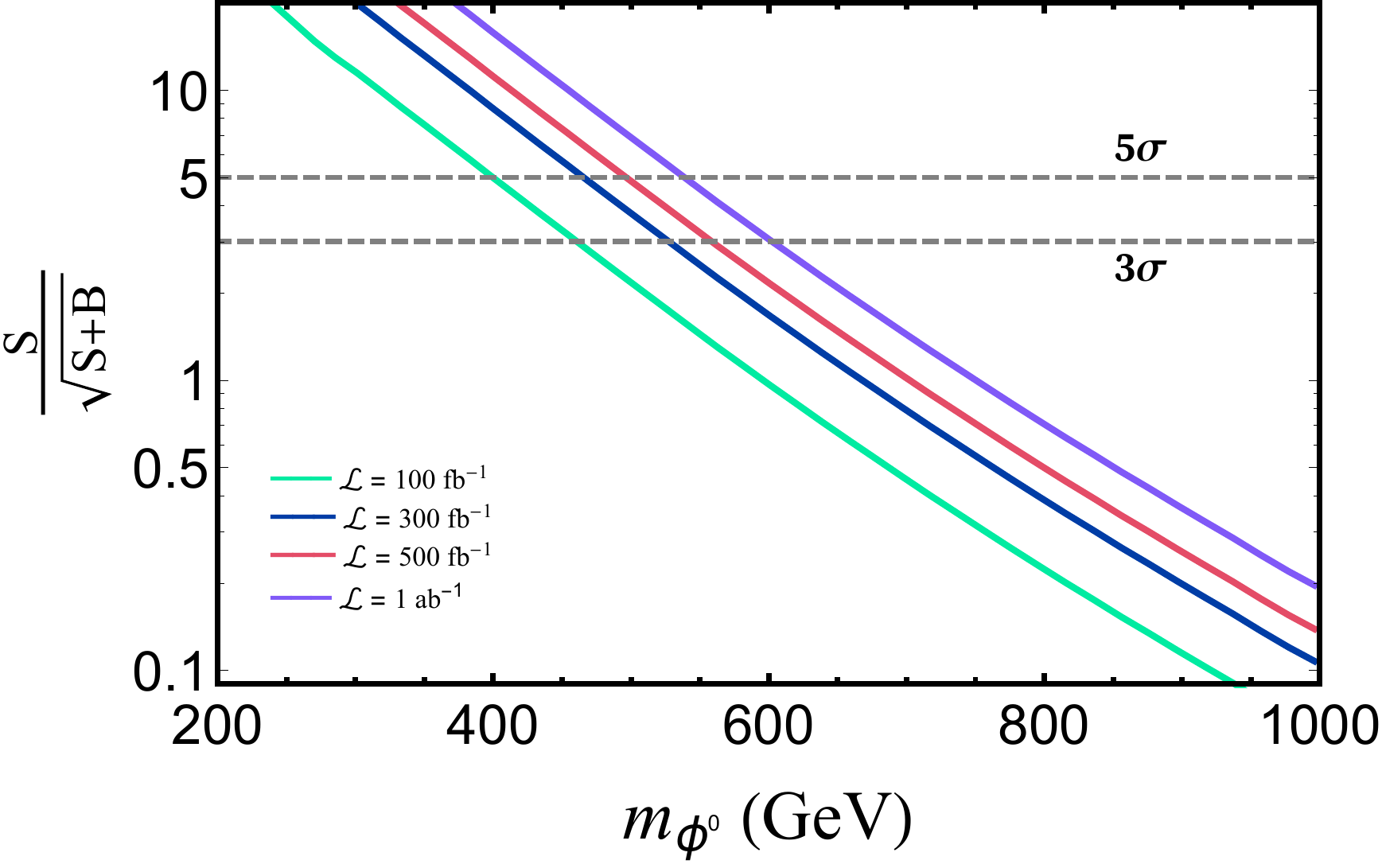}
    \caption{Significance of the signal $pp \to l^+ l^- \tau^+ \tau^-$ in $SU(2)_H$ model   at 14 $\mathrm{TeV}$ LHC  with integrated luminosity  $\mathcal{L}=100,\,300,\,500$ and 1000 fb$^{-1}$. }
    \label{sensitivity_su2h}
\end{figure}

After implementing the $SU(2)_H$ model file in {\tt FeynRules} package~\cite{Christensen:2008py}, we have analyzed the cross section for $p p \to e^- e^+ \tau^- \tau^+, \,\mu^- \mu^+ \tau^- \tau^+$ process  using {\tt MadGraph5aMC@NLO} ~\cite{Alwall:2011uj,Alwall:2014hca}, simulating the hadronization and underlying event effects with {\sc Pythia8}~\cite{Sjostrand:2007gs} and detector effects with the {\sc Delphes3} package~\cite{deFavereau:2013fsa}.  To optimize the signal efficiency over the SM background, we adopt the basic acceptance criteria:  $p_{T}>15 $ {GeV} for each lepton, and  pseudorapidity $|\eta|<2.5 $. In addition, events are required to have exactly two charged lepton candidates with opposite sign and different flavor, i.e., $e\tau$ or $\mu \tau$ and same-flavor lepton pairs are rejected to reduce large SM Drell-Yan background. Although  $\tau$ lepton reconstruction is more challenging amongst the three generation of leptons due to low $\tau$ identification efficiency at the LHC,  it can be reconstructed through their hadronic decays ($\tau_{h}$) \cite{Aad:2019ugc}.  We impose the acceptance cuts for selecting the events with at least 4 leptons with $p_{T} \geq 100,80,40$ and 40 GeV, respectively, and also demand that at least two of them must be $\tau_{\mathrm{h}}$s. In order to minimize the background originating from top decays, we did not allow events with $b$-tagged jets to be accepted. The events with a total transverse hadronic energy (TTHE) greater than $20 ~\mathrm{GeV}$ are also rejected. Finally  a veto to the invariant mass and  transverse mass of the  leptons is implemented compatible with the mass of the $Z$ boson and $W$ boson respectively. We analyze both the signal $p p \to l^- l^+ \tau^-  \tau^+$ and backgrounds. Our results are  summarized in Table.~\ref{tab:lhcana} for five different  masses $m_{\phi_i^0}$= 300, 400, 500, 600 and 800 GeV. After passing through all these 
acceptance criteria, for a 500 GeV massive $\phi_i^0$, we estimate a total of 57 events at $\sqrt{s}$=14 $ \mathrm{TeV}$ LHC with 300 $ \mathrm{fb}^{-1}$ integrated luminosity.  This corresponds to a significance of  $\frac{S}{\sqrt{S+B}}=3.8\sigma$.

In Fig. \ref{sensitivity_su2h}, we estimate the significance of the signal $pp \to l^+ l^- \tau^+ \tau^-$ in our model   at 14 $\mathrm{TeV}$ LHC  with an integrated luminosity of $\mathcal{L}=100,\,300,\,500$ and 1000 fb$^{-1}$. We  find that at 5$\sigma$ level, the scalars $\phi_i^0$ can be probed up to masses of 538, 498, 466 and 398 GeV respectively for the integrated luminosities $\mathcal{L}=100,\,300,500$ and 1000 fb$^{-1}$, whereas at 3$\sigma$ level, this can be probed upto 598, 560, 528 and 458 GeV respectively.
%%%%%%%%%%%%%%%%%%%%%%%%%%%%%%%%%%%%%%%%%
\subsection{Non-Standard interactions and IceCube} 
%%%%%%%%%%%%%%%%%%%%%%%%%%%%%%%%%%%%%%%%%

\begin{figure}[h!]
    \centering
      \includegraphics[width=0.59\textwidth]{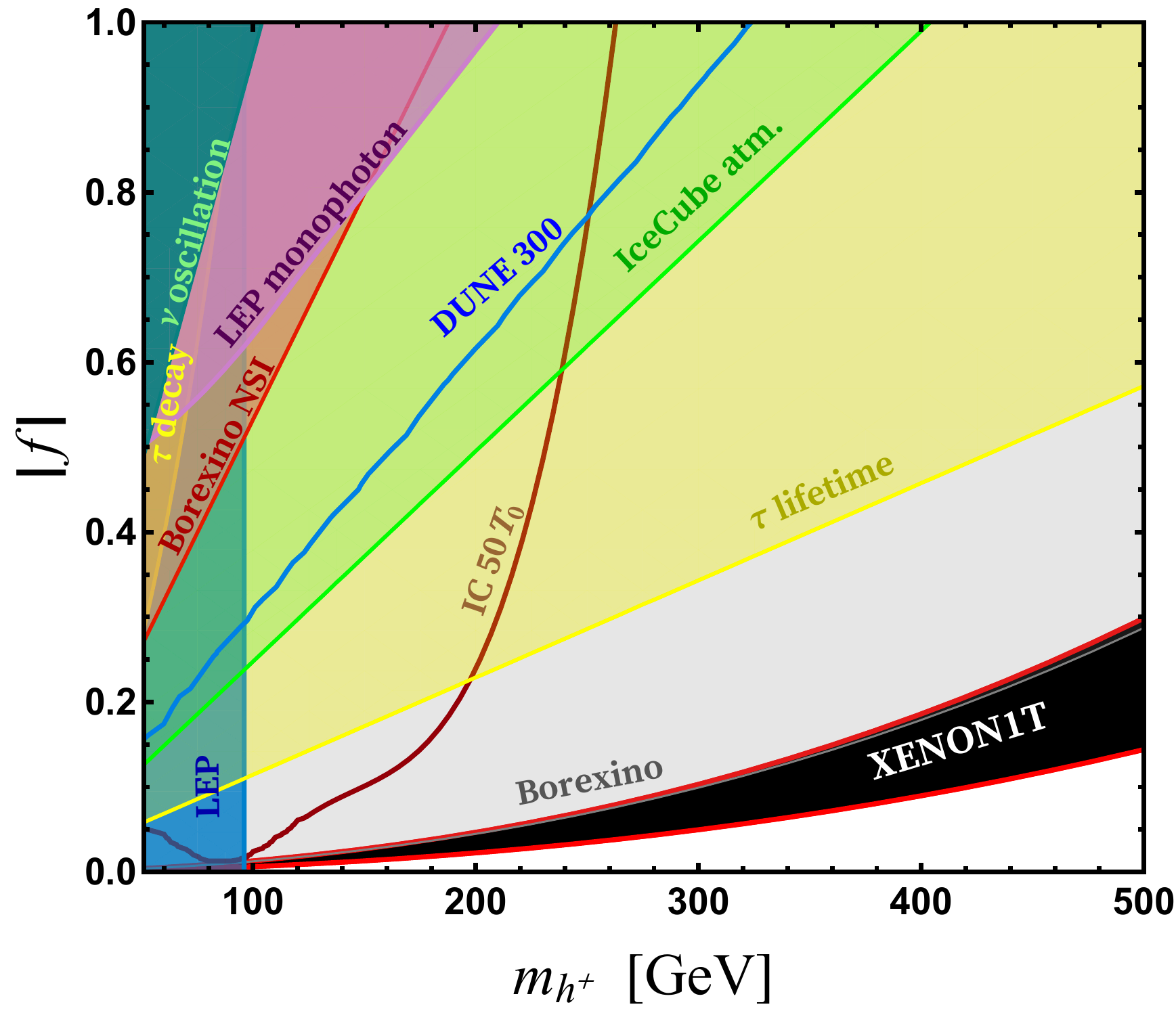}
    \caption{Allowed parameter space of the $SU(2)_H$  model. The region labelled XENON1T explains the electron recoil excess at 90$\%$ C.L.  The blue-shaded region  is excluded by LEP searches of charged scalar; purple-shaded region indicates exclusion from  LEP monophoton search; light yellow-shaded region is excluded from $\tau$ lifetime bound and the dark yellow region is excluded from $\tau $  decay asymmetry constraints. We show the direct constraints on NSI from the global fit to neutrino oscillation + COHERENT data~\cite{Esteban:2018ppq} (dark-green shaded region), neutrino-electron scattering experiments such as Borexino~\cite{Agarwalla:2019smc} (orange shaded region) and IceCube atmospheric neutrino data~\cite{Esmaili:2013fva} (light-green shaded region). The blue solid line indicates the future DUNE sensitivity  for 300 kt.MW.yr exposure~\cite{dev_pondd}. The future IceCube sensitivity is shown by solid brown curve, for an exposure times (50 $\times$ the current exposure $T_{0}=2653$ days). The gray shaded region indicates the current exclusion limit on neutrino magnetic moment from Borexino experiment. Here we set $m_{H^+} = m_{h^+} + 500$ GeV and $f^{\prime}=$1.}
    \label{nsi}
\end{figure}

 The singly-charged scalar $\eta_2^{+}$ can induce non-standard neutrino interaction (NSI) at the tree level via the Yukawa coupling $f$, given by (we use the standard notation, see \cite{Dev:2019anc})
\begin{equation}\varepsilon_{\alpha \beta} \equiv \varepsilon_{\alpha \beta}^{\left(h^{+}\right)}+\varepsilon_{\alpha \beta}^{\left(H^{+}\right)}=\frac{1}{4 \sqrt{2} G_{F}} |f|^2 \left(\frac{\cos ^{2} \alpha}{m_{h^{+}}^{2}}+\frac{\sin ^{2} \alpha}{m_{H^{+}}^{2}}\right)~.
\end{equation}
While significant $\epsilon_{\tau\tau}$ could have been induced within the model, the $\tau $ lifetime constraint restricts   $\epsilon_{\tau \tau} < 2\%$. Also, there are  direct constraints on NSI from neutrino experiments. In Fig.~\ref{nsi}, we have shown these constraints. The best experimental constraint on the NSI parameter $\epsilon_{\tau \tau}$ arises from IceCube atmospheric neutrino data~\cite{Esmaili:2013fva}, which is shown as light-green shaded region in Fig.~\ref{nsi}. We have also included constraints from  global fit to neutrino oscillation + COHERENT data~\cite{Esteban:2018ppq} (dark-green shaded region) and neutrino-electron scattering experiments such as Borexino~\cite{Agarwalla:2019smc}  (orange shaded region). We have also shown future DUNE sensitivity  for 300 kt.MW.yr exposure~\cite{dev_pondd} by blue solid line. The constraint from LEP on charged scalar searches discussed in Sec. \ref{SEC-05} is indicated by blue shaded region.  At LEP there are new contributions to the monophoton process $e^+ e^- \to \nu\bar{\nu}\gamma$ mediated by $\eta_2^+$ in the  $t$-channel, which is bounded from LEP data~\cite{Berezhiani:2001rs}. This limit is shown by the purple shaded region. A light charged scalar $\eta_2^+/\phi_2^+$ of the model could potentially give rise to a Glashow-like resonance feature \cite{Babu:2019vff} in the ultra-high energy neutrino event spectrum at the IceCube; this future IceCube sensitivity is shown by solid brown curve corresponding toan exposure time of 50 $\times T_{0}$. The gray shaded region indicates the current exclusion limit on neutrino magnetic moment from Borexino experiment. One can see that there is a large parameter space in between the two solid red curves which could explain XENON1T electron recoil excess at 90$\%$ C.L. while consistent with all the experimental constraints.

%%%%%%%%%%%%%%%%%%%%%%%%%%%%%%%%%%%%%%%%%%%%%%%
%%%%%%%%%%%%%%%%%%%%%%%%%%%%%%%%%%%%%%%%%%%%%%%
\section{ Generalization to \boldmath$SU(3)_H$ Horizontal Symmetry }\label{SEC-06}
%%%%%%%%%%%%%%%%%%%%%%%%%%%%%%%%%%%%%%%%%%%%%
%%%%%%%%%%%%%%%%%%%%%%%%%%%%%%%%%%%%%%%%%%%%%

In this section we show how the $SU(2)_H$ horizontal symmetry acting on the electron and muon families can be extended to a three-family $SU(3)_H$ symmetry, while preserving the enhancement in neutrino transition magnetic moments relative to their masses.  The main idea is that if the three lepton families transform as a $3$ of an $SU(3)_H$ symmetry, the neutrino magnetic moment term, which is part of the antisymmetric $3^*_a$ in the decomposition $3 \times 3 = 3_a^* + 6_s$ of $SU(3)_H$ may be allowed, while the neutrino mass  term belonging  to the $6_s^*$ could be suppressed. This could happen if the symmetry breaking sector does not include a 6 of $SU(3)_H$, but contains a $3$. We now outline an explicit model that implements this idea.

The electroweak gauge symmetry $SU(2)_L \times U(1)_Y$ of the lepton sector is extended to have a horizontal $SU(3)_H$ symmetry.  The leptons of the SM transofrm under $SU(2)_L \times U(1)_Y \times SU(3)_H$ as follows:
\begin{eqnarray}
        \psi_L &=& \left (\begin{matrix} \nu_e & \nu_\mu & \nu_\tau \\ e & \mu & \tau  \end{matrix}   \right)_L:~~~~ (2, -\frac{1}{2})(3) \nonumber \\
        \psi_R &=& \left(\begin{matrix} e_R~ & \mu_R & \tau_R  \end{matrix}  \right):~~~~~  (2, -1)(3)~.
\end{eqnarray}
New vector-like leptons are introduced to play the role of $\tau$ lepton of $SU(2)_H$:
\begin{eqnarray}
        \chi_{L,R} = \left( \begin{matrix} N \\ E \end{matrix}  \right)_{L,R}: ~~~(2, -\frac{1}{2})(1),~~~~~~E'_{L,R}: ~~~(1,-1)(1)~.
\end{eqnarray}
The Higgs sector consists of triplet fields $\Phi$ and $\eta$ which induce large transition magnetic moments for the neutrino, the SM doublet $\phi_S$, and a flavon field to break $SU(3)_H$ down to $SU(2)_H$:
\begin{eqnarray}
        \Phi(2,\frac{1}{2})(3) &=& \left(\begin{matrix} \phi_1^+ & \phi_2^+ & \phi_3^+ \\ \phi_1^0 & \phi_2^0 & \phi_3^0  \end{matrix}  \right);~~~~~~~\eta(1,1)(3^*) = \left(\begin{matrix}  \eta_1^+ & \eta_2^+ & \eta_3^+ \end{matrix}  \right) \nonumber \\
        \phi_S(2, \frac{1}{2})(1) &=& \left(\begin{matrix}  \phi_S^+ \\ \phi_S^0\end{matrix}   \right),~~~~~~~~~~~~~~~~~~~\varphi(1,1)(3) = \left(\begin{matrix} \varphi_1 & \varphi_2 & \varphi_3 \end{matrix}   \right)~.
\end{eqnarray}
As can be seen, these fields are straightforward generalizations of the fields in the $SU(2)_H$ model, except for the flavon field $\varphi$, which acquires a VEV, $\langle \varphi_3 \rangle = u$, breaking $SU(3)_H$ down to $SU(2)_H$.

\begin{figure}[h!]
    \centering
      \includegraphics[width=0.65\textwidth]{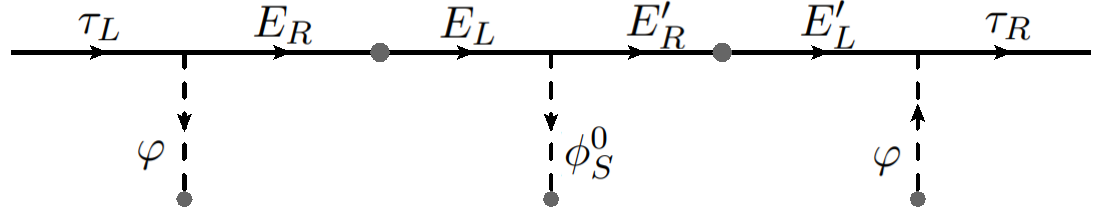}
    \caption{Feynman diagrams generating the tau lepton masses in the $SU(3)_H$ model.}
    \label{su2h_lepmass}
\end{figure}

The $\tau$ lepton would acquire its mass by mixing with the vector-like leptons as shown in Fig. \ref{su2h_lepmass}. The relevant Yukawa Lagrangian invariant under $SU(3)_H$ is given by
\begin{eqnarray}
{\cal L}_{\rm Yuk} &=& y_0 \overline{\psi}_L \psi_R \phi_S + M_E(\overline{N}_L N_R + \overline{E}_LE_R) + M_{E'} \overline{E'}_L E'_R
+ y_1(\overline{N}_L E'_R \phi_S^+ + \overline{E}_L E'_R \phi_S^0) \nonumber \\
&+& y_2 (\overline{N}_R E'_L \phi_S^+ + \overline{E}_R E'_L \phi_S^0) + y_3 \varphi\, \overline{\psi}_L \chi_R + y_4 \varphi\, \overline{\psi}_R E'_L + H.c.       
        \end{eqnarray}

\noindent The $\tau$ lepton mass induced by this Lagrangian can be read off from Fig. \ref{su2h_lepmass}:
\begin{equation}
m_\tau = \left(\frac{y_2 y_3 y_4 u^2}{M_E M_{E'}}\right) \frac{v}{\sqrt{2}}~.
\end{equation}

There are also lepton number violating interactions in the model.  The following additional Yukawa couplings are permitted:
\begin{eqnarray}
    {\cal L}'_{\rm Yuk} &=& {\rm Tr}\left(\psi_L^T i \tau_2 C \chi_L \right) + f' {\rm Tr}\left(\overline{\psi}_L \Phi\right) \nonumber \\
    &=& f\left[\eta_1^+(e^T C N_L - \nu_e^T C E_L) + \eta_2^+(\mu^T C N_L-\nu_\mu^T C E_L) + \eta_3^+(\tau^T C N_L - \nu_\tau^T C E_L)   \right] \nonumber \\
    &+& f' \left[(\overline{\nu}_e \phi_1^+ + \overline{e}_L \phi_1^0) E_R' + (\overline{\nu}_\mu \phi_2^++ \overline{\mu}_L \phi_2^0)E'_R + (\overline{\nu}_\tau \phi_2^+ + \overline{\mu}_L \phi_2^0)E'_R) \right]~.
\end{eqnarray}

The scalar potential contains a term
\begin{eqnarray}
        V &\supset& \lambda_0 \Phi^T_i i\tau_2 \phi_S\eta_j^+ \varphi^T_k \epsilon^{ijk} + H.c. \nonumber \\
        &=& \lambda_0 u \left[(\phi_1^+ \phi_S^0 - \phi_1^0 \phi_s^+) \eta_2^- - (\phi_2^+ \phi_S^0 - \phi_2^0 \phi_S^+) \eta_1^- \right] + H.c.
        \label{cubicsu3}
\end{eqnarray}
Here $\epsilon^{ijk}$ refers to the invariant symbol of $SU(3)_H$, and in the second line, we inserted the VEV of the flavon field $\langle \varphi \rangle = u$, which breaks $SU(3)_H$ down to $SU(2)_H$. Owing to this unbroken symmetry, the masses of $(\phi_1^+,\, \phi_2^+)$ are identical, as are the masses of $(\eta_1^+,\, \eta_2^+)$ fields.  The interactions of Eq. (\ref{cubicsu3}) would mix $(\phi_1^+,\, \eta_2^+)$ and $(\phi_2^+,\,\eta_1^+)$ leading to mass matrices given as
\begin{eqnarray}
        M^2_{\phi_1^+-\eta_2^+} = \left( \begin{matrix} m_\phi^2 & \lambda_0 u v_S \\ \lambda_0 u v_S & m_\eta^2 \end{matrix}   \right),~~~~~    M^2_{\phi_2^+-\eta_1^+} = \left( \begin{matrix} m_\phi^2 & -\lambda_0 u v_S \\ -\lambda_0 u v_S & m_\eta^2 \end{matrix}   \right)
        \label{mixingnew}
\end{eqnarray}
The resulting mass eigenstates $(h_i^+,\, H_i^+)$ for $i=1,2$ have the same mass. However, the mixing angle $\sin2\alpha$ in the two sectors now have an opposite sign.  

\begin{figure}[h!]
    \centering
      \includegraphics[width=\textwidth]{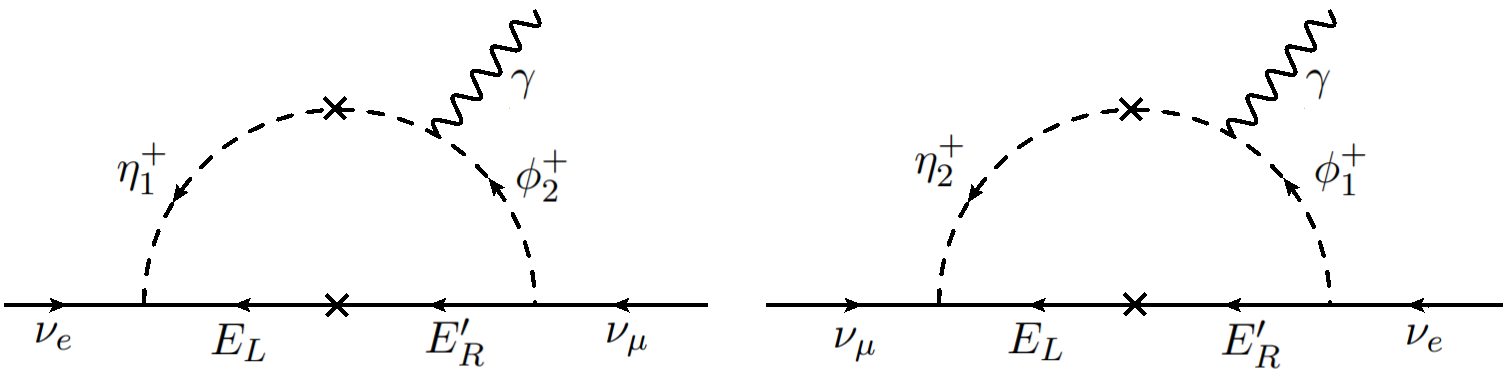}
    \caption{Feynman diagrams generating neutrino transition magnetic moment $\mu_{\nu_e\nu_\mu}$ in the $SU(3)_H$ model. The same diagrams with the photon lines removed could contribute to neutrino mass, but the diagrams cancel in this case.}
    \label{su3h_mag}
\end{figure}

The diagrams shown in Fig. \ref{su3h_mag} will induce an $SU(2)_H$-invariant transition magnetic moment $\mu_{\nu_e\nu_\mu}$.  Owing to the relative minus sign in the mixing angle of Eq. (\ref{mixingnew}), when the photon line is removed in Fig. \ref{su3h_mag}, the two diagram add to yield zero neutrino mass. (All other couplings in the two diagrams are identica.)  For the magnetic moment, the two diagrams add, since $\nu_\mu^T C \sigma_{\mu \nu} \nu_e = -\nu_e^T C \sigma_{\mu \nu} \nu_\mu$.  The resulting $\mu_{\nu_\mu \nu_e}$ is given as in Eq. (\ref{moment}), but with $m_\tau$ replaced by $M_E \xi$, where $\xi$ is a mixing parameter in the $E-E'$ sector, which could be of order $0.1$.  Clearly, large magnetic moment can arise, consistent with neutrino mass as well as other experimental constraints. As for the breaking of the remaining $SU(2)_H$, we adopt the same explicit breaking mechanism of Sec. \ref{SEC-03}.

%{\bf Add a short discussion on the vector-like lepton mass limits here.}

\subsection{Collider signals}

The vector-like leptons ($N, E, E^{\prime}$) present in the $SU(3)_H$ model can be searched for at colliders. Here we briefly highlight their novel signatures and discovery potential at the LHC. We also outline the existing bounds on the masses of these leptons. Although there is mixing between the $SU(2)_L$ doublet lepton $E$ and the singlet lepton $E'$, this mixing is of order 0.1, which we shall ignore for the present discussion.  Being a singlet, $E^{\prime}$ can only be pair-produced ($pp \to E^{\prime^\pm}E^{\prime^\mp}$) via $s-$ channel $Z/\gamma $ exchange, whereas the doublet vector-like lepton $E^\pm$ can also be produced via $s-$ channel $W$ boson exchange:
\begin{equation}
    pp \to E^{\pm}E^{\mp}, \quad pp \to E^{+}N/E^{-}\bar{N} , \quad pp \to N\bar{N}.
\end{equation}
The discovery potential for the doublet vector-like lepton is much brighter than the singlet case, since the largest production rate $ pp \to E^{+}N/E^{-}\bar{N}$ arises from the $s-$ channel $W-$ exchange process. After being produced on-shell, the doublet charged lepton $E^{\pm}$ mostly decays to $Zl^{\pm}$ and $h l^{\pm} $, while the neutral lepton $N$ mostly decays to $W^{\pm} l^{\mp}$. On the other hand, the singlet charged lepton $E^{\prime^\pm}$ has three decay modes to $Zl^{\pm},h l^{\pm}  $ and $W \nu$. Assuming $m_{E^{\prime}} \gg m_{h}, m_{Z}, m_{W},$ the branching ratios to different decay modes  asymptotically approach values given by
\begin{equation}\operatorname{BR}\left(E,E^{\prime} \rightarrow W \nu\right): \operatorname{BR}\left(E,E^{\prime} \rightarrow Z l\right): \operatorname{BR}\left(EE^{\prime} \rightarrow h l\right)=\left\{\begin{array}{ll}
2: 1: 1 & \text { (singlet $E^{\prime}$) } \\
0: 1: 1 & \text { (doublet $E$) }.
\end{array}\right.\end{equation}
The ATLAS Collaboration has searched  for vector-like leptons decaying into  a $Z$ boson and a SM lepton at  center-of-mass energy of $\sqrt{s} = 8 $ TeV, and imposed an exclusion limit  in the mass range of $114-176$ GeV from non-observation of the signal events \cite{Aad:2015dha}.  LEP experiments have imposed a lower limit on vector-like lepton mass  of $\approx 100 $ GeV \cite{Achard:2001qw}. Recently, CMS collaboration has performed a search for vector like leptons \cite{Sirunyan:2019ofn} looking at multi-lepton final states at center-of-mass energy of $\sqrt{s} = 13 $ TeV with integrated luminosity
 $\mathcal{L}=77.4 $ fb$^{-1}$ and  imposed the best limit on vector-like lepton doublet mass up to 790 GeV at $95\%$ C.L.  Similar phenomenological implications of vector-like leptons can be found in Ref.  \cite{Kumar:2015tna,Falkowski:2013jya}, although in a different context.
%\begin{figure}[h!]
 %   \centering
  %    \includegraphics[width=0.5\textwidth]{col2.png}
   % \caption{Feynman diagram for the signature $p p \to l^- l^+ + {E\!\!\!\!/}_{T} $ in the $SU(3)_H$ model at the LHC via exchange of the exotic charged leptons.}
   % \label{col2}
%\end{figure}

%%%%%%%%%%%%%%%%%%%%%%%%%%%%%%%%%%%%%%%%%%%%%%%
%%%%%%%%%%%%%%%%%%%%%%%%%%%%%%%%%%%%%%%%%%%%%%%
\section{ Neutrino Magnetic Moment in the Zee model }\label{SEC-07}
%%%%%%%%%%%%%%%%%%%%%%%%%%%%%%%%%%%%%%%%%%%%%
%%%%%%%%%%%%%%%%%%%%%%%%%%%%%%%%%%%%%%%%%%%%%

In 1990,  Barr, Freire, and Zee (BFZ) proposed a spin symmetry mechanism \cite{Barr:1990um} (reviewed briefly in Sec. 3) which provides for a large neutrino transitional magnetic moment $\mu_{\nu}$, with a relatively small neutrino mass. To illustrate the mechanism, they extended the scalar sector of the popular Zee model of neutrino mass \cite{Zee:1980ai} with an additional Higgs  doublet. Subsequently it was shown in Ref. \cite{Babu:1992vq}  that this mechanism can be realized within the Zee model without the addition of a third scalar doublet, providing large neutrino magnetic moment.  However, the contribution of two-loop graphs for the neutrino transition magnetic moments have not been quantitatively analyzed thus far. Here we perform such an analysis and derive admissible values of the neutrino transition magnetic moment in the Zee model as well as in its BFZ extension in light of the current constraints from colliders as well as from low energy experiments.
%\subsection{Model description }

The simplest realization of the spin-symmetry mechanism is  the Zee model~\cite{Zee:1980ai}, which contains an $SU(2)_L$-doublet scalar $H_2$  and an $SU(2)_L$-singlet charged scalar $\eta^\pm$, in addition to the SM-like Higgs doublet $H_1$. The Wolfenstein version of the  model~\cite{Wolfenstein:1980sy}, which is more predictive by virtue of a $Z_2$ symmetry  is ruled out by oscillation data~\cite{Koide:2001xy, He:2003ih}. However, it has been shown that the original version of the Zee model~\cite{Zee:1980ai} is fully consistent with neutrino oscillation data with interesting phenomenology \cite{Herrero-Garcia:2017xdu, Babu:2019mfe}.

Here, we mainly concentrate on the prediction of the neutrino transition magnetic moment in the Zee model. We adopt the scalar potential and the resulting scalar mass spectrum and the conventions of Ref. \cite{Babu:2019mfe}. We choose a rotated basis for the Higgs doublets \cite{Babu:2018uik} in which only one neutral Higgs $H_1$ has a nonzero vacuum expectation value.  Specifically,
\begin{equation}
H_{1}=\left(\begin{array}{c}
G^{+} \\
\frac{1}{\sqrt{2}}\left(v+\phi_{1}^{0}+i G^{0}\right)
\end{array}\right), \quad H_{2}=\left(\begin{array}{c}
H^{+} \\
\frac{1}{\sqrt{2}}\left(\phi_{2}^{0}+i A\right)
\end{array}\right),
\end{equation}
where $G^{+}$ and $G^{0}$ denote unphysical Goldstone bosons, $H^{+}$ is the physical charged Higgs boson, $\phi_{1}^{0}, \phi_{2}^{0}$ represent  CP-even neutral Higgs fields (not the mass eigenstates) and $A$ is a CP-odd neutral Higgs field. The VEV is defined as $v=\sqrt{v_{1}^{2}+v_{2}^{2}} \simeq$ 246 GeV.  The physical scalar spectrum contains three neutral scalars $\varphi_{j}^{0}=\{h, H, A\},$ which are related with the original  neutral fields via an orthogonal transformation $\varphi_{j}^{0}=\mathcal{R}_{j i} \phi_{i}.$ Similarly, the two mass eigenstate for the charged scalars are related   with the original fields via a rotation matrix $\mathcal{\zeta}_{i j}$ as follows:
%\begin{equation}
%\left(\begin{array}{c}
%h\\
%H \\
%A
%\end{array}\right)
%=\mathcal{R}\left(\begin{array}{c}
%\phi_{1}^0 \\
%\phi_{2}^0 \\
%\phi_{3}^0
%\end{array}\right), \end{equation}
\begin{equation}\left(\begin{array}{c}
h^+ \\
H^+
\end{array}\right)=\zeta \left(\begin{array}{c}
\eta^{+} \\
H_{2}^+~.
\end{array}\right)\end{equation}
The Yukawa Lagrangian can be written as:  
\begin{equation}\begin{aligned}
\mathcal{L}_{y} &=Y_{d} \bar{Q}_{L} d_{R} H_{1}+\widetilde{Y}_{d} \bar{Q}_{L} d_{R} H_{2}+Y_{u} \bar{Q}_{L} u_{R} \widetilde{H}_{1}+\widetilde{Y}_{u} \bar{Q}_{L} u_{R} \widetilde{H}_{2} \\
&+Y_{\ell} \bar{\psi}_{L} H_{1} \psi_{R}+\widetilde{Y}_{\ell} \bar{\psi}_{L} H_{2} \psi_{R}+ f\bar{\psi}_{L} {\psi}_{L} \eta^+ +\text H.c.
\end{aligned}\end{equation}
where  $\psi_{\ell}=(\nu, e)_{L}^{T}$ and $Q_{L}=(u, d)_{L}^{T}$ represent the left-handed lepton and quark doublets, $f$ is an antisymmetric Yukawa coupling matrix in flavor space ($\left.f_{\alpha \beta}=-f_{\beta \alpha}\right),$ and  $Y$ and $\widetilde{Y} $ are  general complex Yukawa matrices.  Since the VEV of $H_{2}$ is zero, the quark and charged lepton mass matrices are given by
\begin{equation}
M_{u}=Y_{u} v / \sqrt{2}, \quad M_{d}=Y_{d} v / \sqrt{2}, \quad M_{l}=Y_{\ell} v / \sqrt{2}~.
\end{equation}

The scalar potential contains a cubic coupling given by
\begin{equation}
    V \supset \mu H_1^i H_2^j \eta^- \epsilon_{ij} + H.c.
\end{equation}
which leads to mixing between $H_2^+$ and $\eta^+$, with a mixing angle denoted as $\varphi$.  Neutrino masses are generated at one-loop level and are given by \cite{Zee:1980ai}
    \begin{equation}
         M_\nu \ = \ \kappa \, (f M_\ell Y + \widetilde{Y}^T M_\ell f^T) \, , \quad \kappa \ = \ \frac{1}{16 \pi^2} \sin{2 \varphi} \log\left(\frac{m_{h^+}^2}{m_{H^+}^2}\right) \, .
         \label{nuMass}
    \end{equation}

%\subsection{Neutrino masses and magnetic moments }
\begin{figure}[h!]
\centering
    \includegraphics[width=0.65\textwidth]{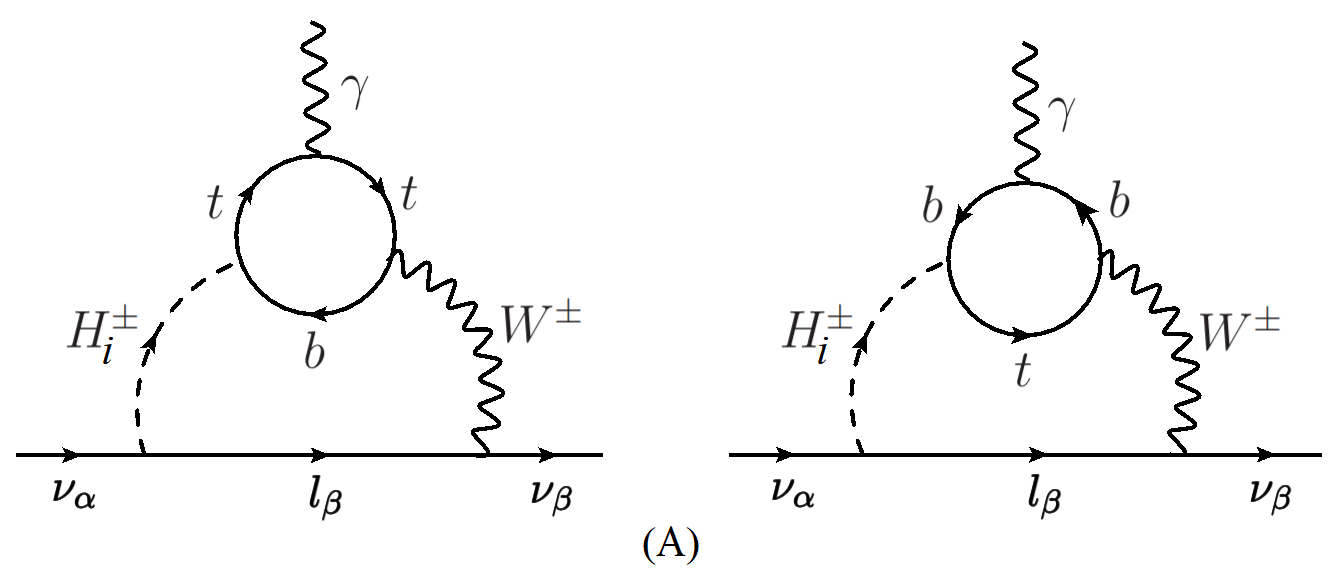}
    \includegraphics[width=0.65\textwidth]{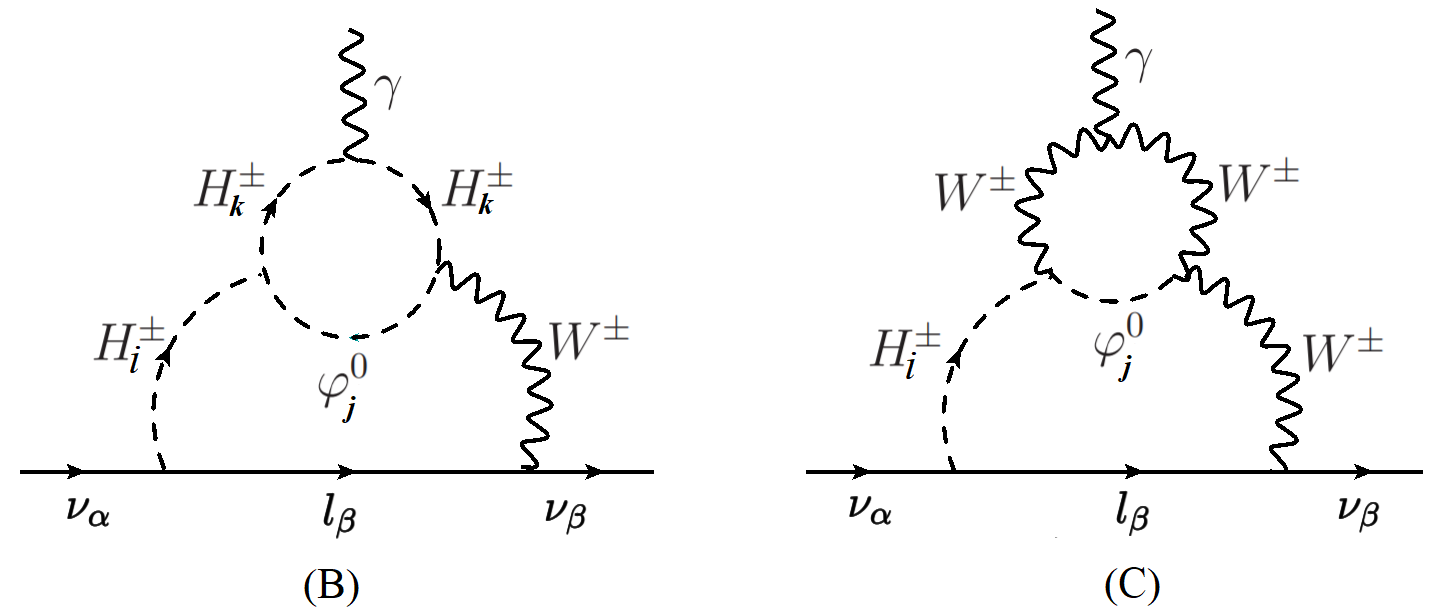}
    \caption{Feynman diagrams for the neutrino transitional magnetic moment, via the quark Yukawa coupling and via the Higgs quartic coupling. We categorize three contributions like this: (A) $q\bar{q} \gamma$ vertex: the contribution from the two diagrams in the top panel , (B) $H_k^+ H_k^- \gamma$ vertex: contribution from the left figure in the bottom panel and (C) $W^+ W^- \gamma$ vertex: contribution from the right figure in the bottom panel.    }
    \label{feynman_spin1}
\end{figure}
Here we will analyze the neutrino transition magnetic moment and its relation to the neutrino mass in the Zee model. In Fig.~\ref{feynman_spin1}, we show all the leading Barr-Zee \cite{Barr:1990vd} diagrams that contribute to a large neutrino magnetic moment. It is these diagrams that enhance $\mu_{\nu_e\nu_\mu}$, as the mass contributions obtained from the same diagrams with the photon lines removed would have an additional suppression factor of $(m_\mu^2/m_W^2)$.  As noted in Sec. 3, with this suppression, $m_\nu \sim 0.1$ eV, when $\mu_\nu \sim 10^{-11} \mu_B$ is realized.  Note that $f_{e\tau}$ cannot contribute to a large $\mu_{\nu_e\nu_\tau}$ via such diagrams, as the resulting mass contribution would be $0.1~{\rm eV} (m_\tau^2/m_\mu^2) \sim 25$ eV, which is excessive.  We thus focus on $\mu_{\nu_e\nu_\mu}$ induced via $f_{e\mu}$.

We have generalized the calculation of muon $g-2$ in two Higgs doublet model of Ref. \cite{Ilisie:2015tra,Abe:2013qla} to the magnetic moment of the muon in the Zee model.
$\mu_{\nu_\mu \nu_e}$ arising from the three sets of diagrams in Fig. \ref{feynman_spin1} are  found to be:
\begin{equation}\label{maga}\begin{aligned}
\mu_{\nu_e \nu_{\mu}}^{(A)}=& \frac{2 \alpha m_{e} N_{C}\left|V_{t b}\right|^{2}}{32 \pi^{3} s_{\mathrm{w}}^{2} } \mathlarger{\mathlarger{\sum}}_{H_i^+=h^+, H^+}^{} \frac{\mathcal{\zeta}_{i 1} \mathcal{\zeta}_{i 2}}{\left(m_{H_i^{\pm}}^{2}-m_{W}^{2}\right)}\int_{0}^{1} d x\left[Q_{t} x+Q_{b}(1-x)\right]\times\left[\operatorname{Re}\left(\widetilde{Y}_{d} f_{e \mu}^{*}\right) \right. \\
& \left. m_b x(1-x) +\operatorname{Re}\left(\widetilde{Y}_{u} f_{e \mu}^{*}\right) m_{t} x(1+x)\right]\left[\mathcal{G}\left(\frac{m_{t}^{2}}{m_{H_i^{\pm}}^{2}}, \frac{m_{b}^{2}}{m_{H_i^{\pm}}^{2}}\right)-\mathcal{G}\left(\frac{m_{t}^{2}}{m_{W}^{2}}, \frac{m_{b}^{2}}{m_{W}^{2}}\right)\right]
\end{aligned}\end{equation}

\begin{equation}\label{magb}\begin{aligned}
\mu_{\nu_e \nu_{\mu}}^{(B)}=&  \frac{2 \alpha m_{e} v }{64 \pi^{3} s_{\mathrm{w}}^{2} } \mathlarger{\mathlarger{\sum}}_{H_i^+=h^+, H^+ ~}^{} \mathlarger{\mathlarger{\sum}}_{\varphi_j^0=h, H, A~ ~}^{}  \mathlarger{\mathlarger{\sum}}_{H_k^+=h^+, H^+ ~ }^{} \frac{\mathcal{\zeta}_{i 1} }{\left(m_{H_i^{\pm}}^{2}-m_{W}^{2}\right)} \operatorname{Re}\left[f_{e \mu}^{*}\left(\mathcal{R}_{j 2}-i \mathcal{R}_{j 3}\right)\right] \\ & \lambda_{\varphi_{j}^{0} H_i^{+} H_k^{-}} \int_{0}^{1} d x x^{2}(x-1)  
 \times\left[\mathcal{G}\left(\frac{m_{H_k^{\pm}}^{2}}{m_{H_i^{\pm}}^{2}}, \frac{m_{\varphi_{j}}^{2}}{m_{H_i^{\pm}}^{2}}\right)-\mathcal{G}\left(\frac{m_{H_k^{\pm}}^{2}}{m_{W}^{2}}, \frac{m_{\varphi_{j}^{0}}^{2}}{m_{W}^{2}}\right)\right]
\end{aligned}\end{equation}

\begin{equation}\label{magc}\begin{aligned}
\mu_{\nu_e \nu_{\mu}}^{(C)}=&  \frac{2 \alpha m_{e}  }{64 \pi^{3} s_{\mathrm{w}}^{2} v } \mathlarger{\mathlarger{\sum}}_{H_i^+=h^+, H^+ ~}^{} \mathlarger{\mathlarger{\sum}}_{\varphi_j^0=h, H, A~ ~}^{}   \frac{1}{\left(m_{H_i^{\pm}}^{2}-m_{W}^{2}\right)} \operatorname{Re}\left[f_{e \mu}^{*} \mathcal{R}_{j 1}\left(\mathcal{R}_{j 2}-i \mathcal{R}_{j 3}\right)\right] \int_{0}^{1} d x x^{2}  \\&
\times \left[\left(m_{H_i^{\pm}}^{2}+m_{W}^{2}-m_{\varphi_{j}^{0}}^{2}\right)(1-x)-4 m_{W}^{2}\right]\left[\mathcal{G}\left(\frac{m_{W}^{2}}{m_{H_i^{\pm}}^{2}}, \frac{m_{\varphi_{j}^{0}}^{2}}{m_{H_i^{\pm}}^{2}}\right)-\mathcal{G}\left(1, \frac{m_{\varphi_{j}^{0}}^{2}}{m_{W}^{2}}\right)\right]
\end{aligned}\end{equation}
where
\begin{equation}
\mathcal{G}\left(\omega^{a}, \omega^{b}\right)=\frac{\ln \left(\frac{\omega^{a} x+\omega^{b}(1-x)}{x(1-x)}\right)}{x(1-x)-\omega^{a} x-\omega^{b}(1-x)}~.
\end{equation}
In Zee model, the cubic scalar coupling  $ \lambda_{\varphi_{j}^{0} H_i^{+} H_k^{-}}$ can be written as
\begin{equation}
 \lambda_{\varphi_{j}^{0} H_i^{+} H_k^{-}} = \lambda_7 \mathcal{R}_{j 2}\mathcal{\zeta}_{i 2}\mathcal{\zeta}_{k 2}  + \frac{\mu}{v} \mathcal{R}_{j 1} \mathcal{\zeta}_{i 1} \mathcal{\zeta}_{k 2} ~.
 \label{cubic3}
\end{equation}

By analyzing the contributions from the diagrams of Fig. \ref{feynman_spin1}, we find that one can achieve neutrino transition magnetic moment as big as $\mu_{\nu_e \nu_\mu} \simeq 3 \times 10^{-12} \mu_B$, which is not sufficient to explain the observed XENON1T electron recoil excess \cite{Aprile:2020tmw}. Now we shall explain the strategy we adopted for the optimization of $\mu_{\nu_e\nu_\mu}$ in the model.
There is stringent constraint on the Yukawa coupling $f_{e\mu}$ from lepton/hadron universality  \cite{Herrero-Garcia:2014hfa}, which can be translated to $f_{e \mu} <\frac{1.674 \times 10^{-4}}{\sqrt{\frac{\cos^2{\varphi}}{m_{h^{+}}^{2}}+\frac{\sin^2{\varphi}}{m_{H^{+}}^{2}}}}$, where $m_{h^{+}}$ and $m_{H^{+}}$ are expressed in GeV. For our analysis we allow the maximum value of $f_{e \mu}$ consitent with this constraint.  Now, in order to get the maximum value for the magnetic moment, one has to set the value of $m_{H^{+}}$  as low as possible. To be consistent with electroweak $T$ parameter constraint, we cannot split the masses among the $H_2$ multiplet by too much, we choose the splittings to be $< \mathcal{O}$(100 GeV) between charged Higgs and neutral Higgs from $H_2$. The LEP experiments exclude charged Higgs mass below $100$ GeV \cite{Babu:2019mfe} from direct searches. There will be other collider consequences which we shall elaborate on  now. 

\begin{figure}[h!]
    \centering
      \includegraphics[width=.6\textwidth]{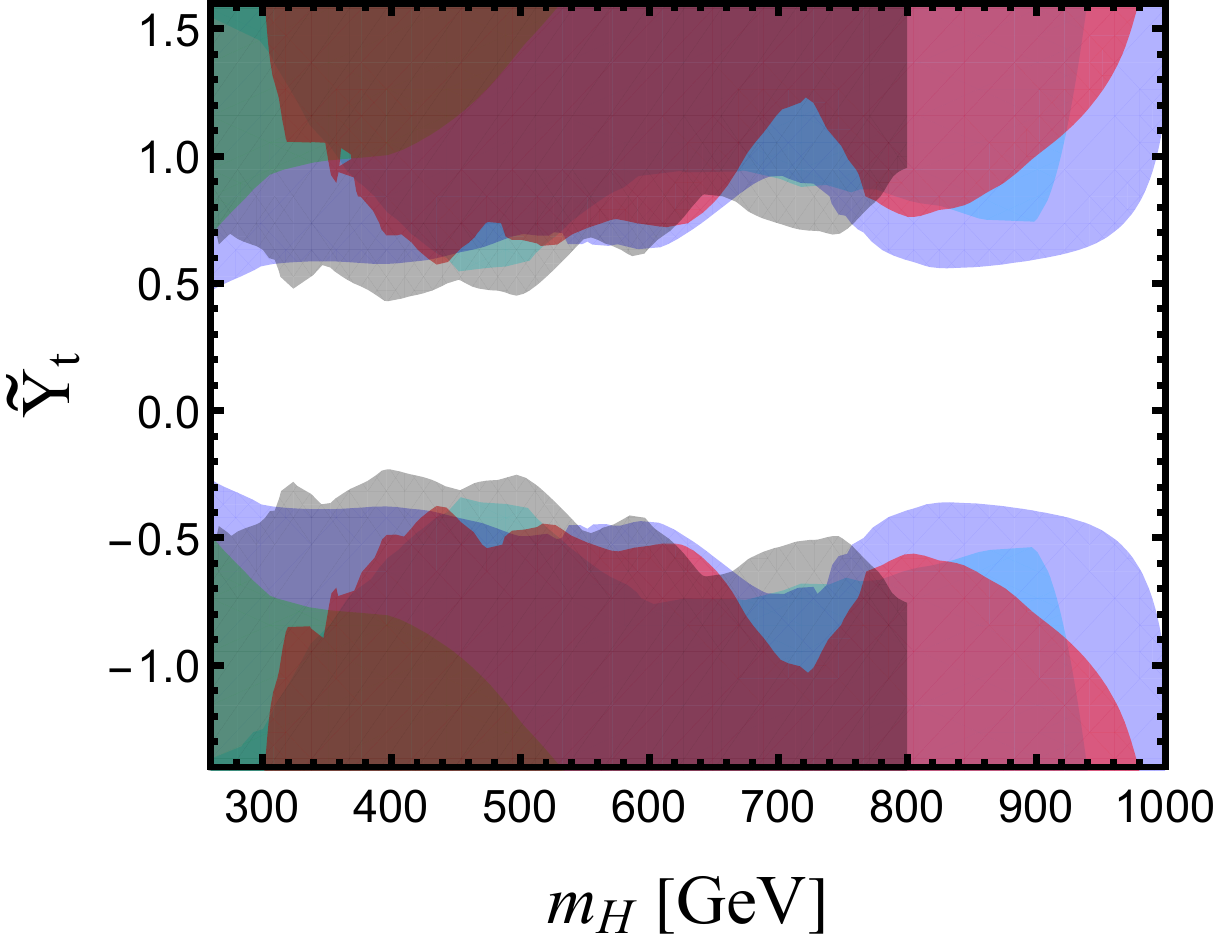}
    \caption{Current limits in top quark Yukawa coupling in Zee model  in $ \tilde{Y}_{t} - m_{H}$ plane from SM Higgs observables as well as Heavy Higgs searches.  Gray, red and cyan shaded regions are excluded from current di-Higgs limit looking at different final states $b \bar{b} \gamma \gamma$ \cite{Aaboud:2018ftw}, $b \bar{b} b \bar{b}$ \cite{Aaboud:2018knk} and $b \bar{b} \tau^{+} \tau^{-}$ \cite{Aaboud:2018sfw} respectively; Blue and green shaded zones are excluded  from the resonant $Z Z$ and $W^{+} W^{-}$ searches \cite{ATLAS:2017spa,CMS:2017skt}. }
    \label{zeebound}
\end{figure}

The top-bottom loop contribution is numerically larger than the scalar loop contribution in Fig. \ref{feynman_spin1}, owing to a color factor and an extra factor of 2 arising from Dirac trace, so we focus on this contribution first. From Eq.~(\ref{maga}), we see that the contribution proportional to the top mass will dominate, which has a  linear dependence on  $\widetilde{Y}_t$. Thus, one has to set $\widetilde{Y}_t$ as large as possible, while being consistent with perturbativity and  other experimental constraints.  Now it turns out that $\widetilde{Y}_t$ is tightly constrained from the searches of SM Higgs observables \cite{Aad:2019mbh} at the LHC as well as from heavy Higgs searches \cite{Aaboud:2018ftw,Aaboud:2018knk,Aaboud:2018sfw,ATLAS:2017spa,CMS:2017skt, Sirunyan:2018wnk}. We summarize all these existing collider bounds  in $ \tilde{Y}_{t} - m_{H}$ plane in Fig.~\ref{zeebound}.  Gray, red and cyan shaded regions are excluded from current di-Higgs limits with final states  $b \bar{b} \gamma \gamma$ \cite{Aaboud:2018ftw}, $b \bar{b} b \bar{b}$ \cite{Aaboud:2018knk} and $b \bar{b} \tau^{+} \tau^{-}$ \cite{Aaboud:2018sfw} respectively. Blue and green shaded zones are excluded  from the resonant $Z Z$ and $W^{+} W^{-}$ searches \cite{ATLAS:2017spa,CMS:2017skt}.  As we can see, for a heavy Higgs mass of 260 GeV, we can allow $\widetilde{Y}_t$ to be at most 0.5. On the other hand, if we go below 250 GeV mass, then on-shell di-Higgs production will be turned off and hence most of the parameter space will be ruled out from non-observation of di-photonss, $WW$ or $ZZ$ resonances\cite{ATLAS:2017spa,CMS:2017skt, Sirunyan:2018wnk}. Thus, we set $m_H $ to be 260 GeV and $\widetilde{Y}_t$  to be 0.5.   

\begin{figure}[t!]
\centering
    \includegraphics[width=0.45\textwidth]{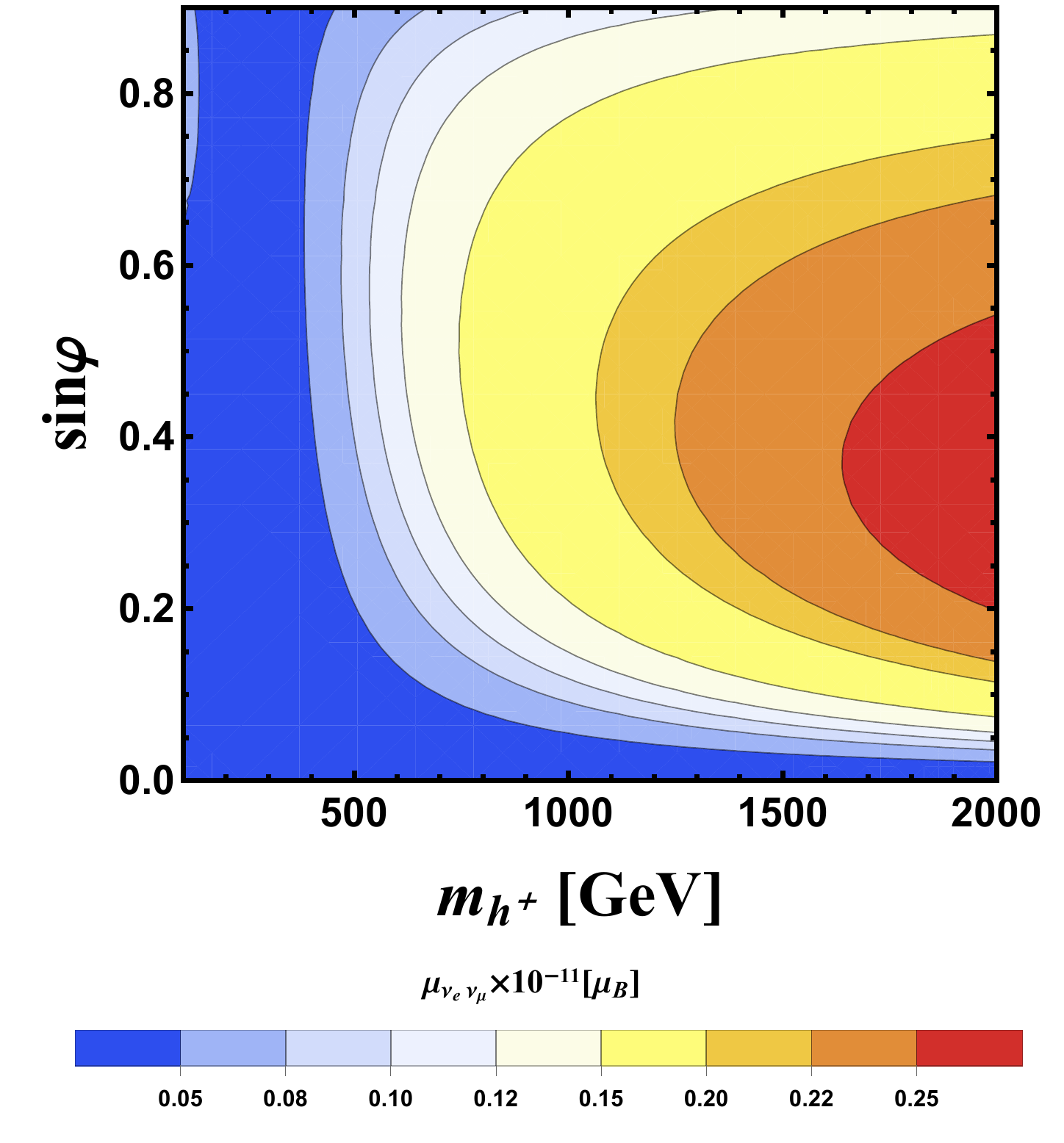}
     \includegraphics[width=0.45\textwidth]{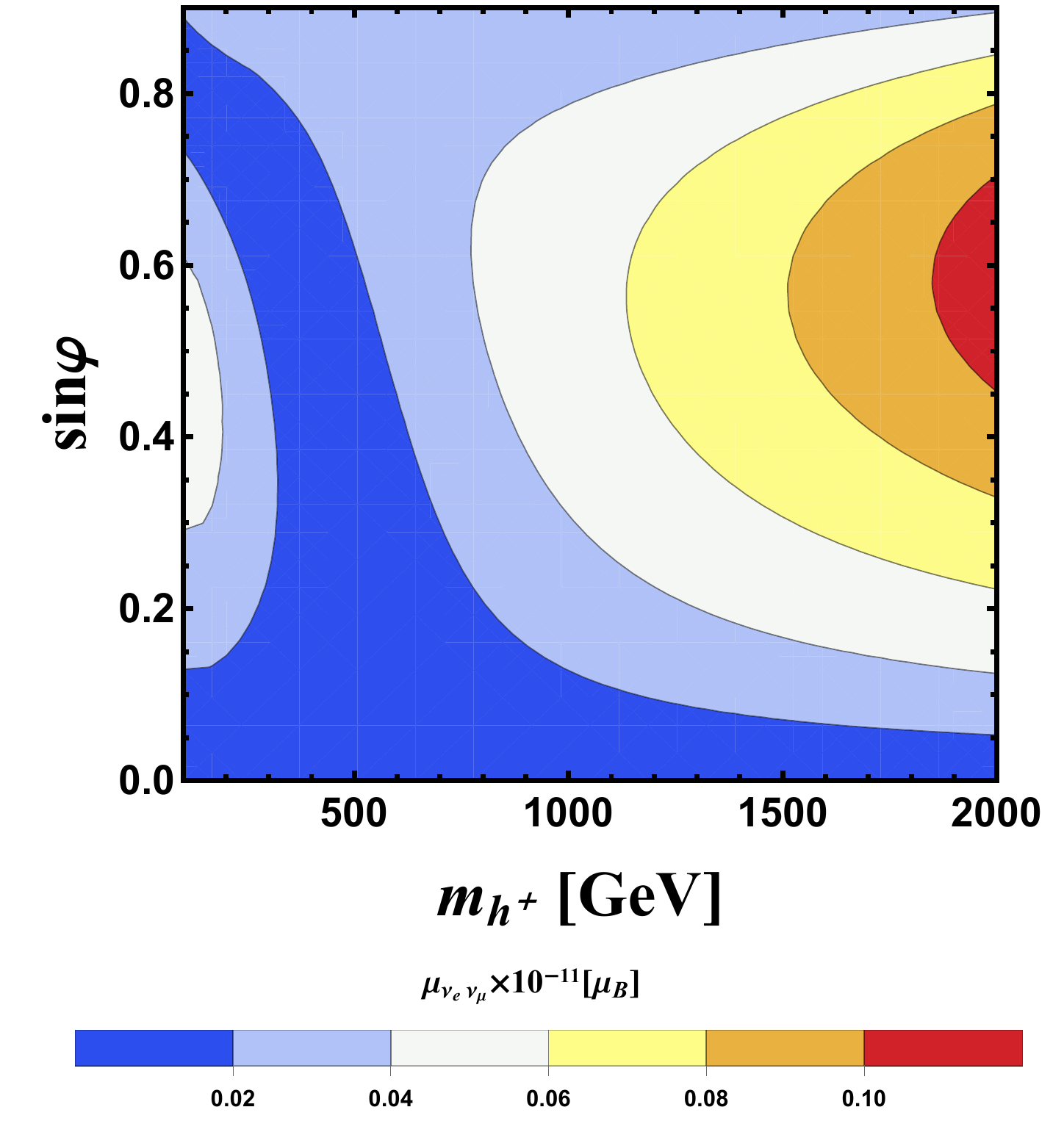}
    \caption{Prediction of neutrino magnetic moment in $m_h^+ - \sin{\varphi}$ plane in the Zee model. Left panel: contribution from the top-bottom loop; right panel: through Higgs cubic and quartic couplings.}
    \label{nmm_zee}
    \end{figure}

We have evaluated the neutrino transition magnetic moment as a function of $h^+$ mass and the mixing angle $\varphi$ in the Zee model.  Our results are shown   in Fig.~\ref{nmm_zee} in the  $m_h^+ - \sin{\varphi}$ plane. The left panel represents the prediction of neutrino transition magnetic moment via top-bottom loop contribution and the right panel is the contribution from the scalar loops. The cubic coupling $\mu$ is automatically  fixed for each point in Fig.~\ref{nmm_zee} while fixing $m_{h^+}$ and  $\sin{\varphi}$.  This is because the mixing angle $\varphi$ and the two charged scalar masses will fix the $\mu$ term. From Fig.~\ref{nmm_zee} we see that the contribution from the scalar loops is suppressed compared to the top-bottom loop contribution by a factor of 4 or so. In this optimized setup, one can achieve neutrino transition magnetic moment  as big as $\mu_{\nu_e \nu_\mu} \simeq 3 \times 10^{-12} \mu_B $, which  is insufficient to explain the observed XENON1T electron recoil excess \cite{Aprile:2020tmw}. We also observe that our analysis is equally applicable to predictions of neutrino transition magnetic moment in extensions of the Zee model making use of the spin-symmetry mechanism. We have extended our analysis of  $\mu_{\nu_e \nu_\mu}$ to the  BFZ  model \cite{Barr:1990um}. Due to the presence of an extra scalar doublet there, the cubic scalar coupling is free compared to the Zee model, see Eq. (\ref{cubic3}). However this cubic coupling in the BFZ model is bounded from unitarity constraints \cite{Goodsell:2018tti} and we can gain a factor  2 to 3  from here compared to the scalar loop contribution of the Zee model, so that $\mu_{\nu_e\nu_\mu} \sim 3 \times 10^{-12} \mu_B$ may be obtained. This is however not sufficient to achieve the desired values to explain the observed XENON1T electron recoil excess.    

\section{ Mechanism to evade astrophysical limits on neutrino magnetic moments}\label{SEC-08}

As noted in Sec. \ref{SEC-03}, Majorana neutrino transition magnetic moments may be severely constrained by stellar energy loss arguments \cite{Bernstein:1963qh,Raffelt:1999tx}.  Photons inside the stars, which has a plasma mass, can decay into neutrinos that would escape, thus contradicting the successful stellar evolution models. The red giant branch of globular clusters provides the most stringent limits, $\mu_\nu < 4.5 \times 10^{-12} \mu_B$ \cite{Viaux:2013lha}, which is in conflict with  the value of $\mu_{\nu_e\nu_\mu} \in(1.65 - 3.42) \times 10^{-11} \mu_B$ that is needed to explain the XENON1T excess.  Here we provide a mechanism that evades this astrophysical bound on $\mu_\nu$ by invoking new interactions of the neutrino with a light scalar.  In the presence of such interactions, neutrinos would acquire a medium-dependent mass, which may exceed the core temperature of the star, thus preventing  plasmon decay kinematically. 

We shall closely follow the recent field theoretic evaluation of the medium-dependent mass of the neutrino in the presence of a  light scalar that also couples to ordinary matter \cite{Babu:2019iml} in illustrating our mechanism. This work follows the observation that such interactions would provide the neutrino with a matter-dependent mass \cite{Ge:2018uhz}.  Phenomenological implications of this scenario, including long-range force effects, were studied in Ref. \cite{Smirnov:2019cae}.  Ref. \cite{Babu:2019iml} analyzed phenomenological constraints from laboratory experiments, fifth force experiments, astrophysics and cosmology.  We shall make use of these constraints here in providing a neutrino trapping mechanism.

\begin{figure}[h!]
    \centering
      \includegraphics[width=.31\textwidth]{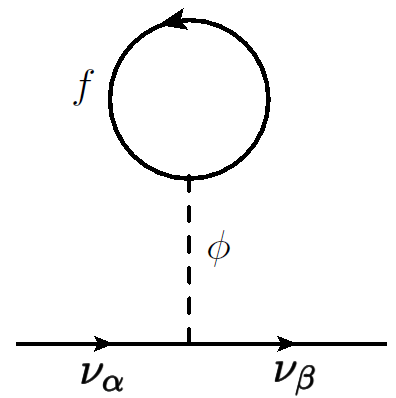}
    \caption{Neutrino self-energy diagram in the background of fermion $f = e$ or $N$, the nucleon. }
    \label{tadpole}
\end{figure}

Consider the interactions of the three Majorana neutrinos $\nu_\alpha$ with a light scalar, which also couples to fermion $f$ which is either the electron or the nucleon: 
\begin{equation}
%\label{eq:LagMN}
    \mathcal{L} \ \supset \ - \frac{y_{\alpha\beta}}{2} \overline{\nu^c_\alpha} \phi\nu_\beta - y_f \bar{f}\phi f -  \frac{m_{\alpha\beta}}{2} \overline{\nu^c_\alpha} \nu_\beta - \frac{m_\phi^2}{2}\phi^2~.
    \label{Majorana}
\end{equation}
These interactions induce a finite density neutrino mass through the diagram shown in Fig. \ref{tadpole}.  Using quantum field theory at finite temperature and density the induced neutrino mass arising from such diagrams has been computed to be \cite{Babu:2019iml}
\begin{align}
  \Delta m_{\nu, {\alpha\beta}} \ = \  \frac{y_{\alpha\beta}y_fm_f}{\pi^2m_\phi^2}\int_{m_f}^\infty dk_0\:\sqrt{k_0^2-m_f^2} \left[n_f(k_0)+n_{\bar f}(k_0)\right]  \ ~.
    \label{eq:sNSI}
\end{align}
Here $n_f$ and $n_{\bar{f}}$ are the occupations numbers of the background fermions and antifermions. This integral can be evaluated analytically in several interesting regimes:
\begin{numcases}{\Delta m_{\nu, {\alpha\beta}} \ = \ }
\frac{y_f y_{\alpha\beta}}{m_\phi^2}\left(N_f+N_{\bar f}\right) & \textbf{     ($\mu, T\ll m_f$)} \nonumber  \\
 \frac{y_f y_{\alpha\beta}}{m_\phi^2}\frac{m_f}{2}\left(\frac{3}{\pi}\right)^\frac{2}{3} \left(N_f^{2/3}+N_{\bar f}^{2/3}\right) & \textbf{($\mu>m_f\gg T$)}  \nonumber\\
 \frac{y_f y_{\alpha\beta} m_f}{3 \:m_\phi^2}\left(\frac{\pi^2}{12 \:\zeta(3)}\right)^\frac{2}{3} \left(N_f^{2/3}+N_{\bar f}^{2/3}\right) & \textbf{ ($\mu<m_f\ll T$)}~. \label{eq:c3}
\end{numcases}
The nonrelativistic low temperature expansion is valid for both the electron and the nucleon in red giant stars ($T \simeq 10$ keV), while the high chemical potential expansion is valid for electron background in supernovae which has $\mu \simeq 150~{\rm MeV} >> m_e$.  The last expansion in Eq. (\ref{eq:c3}) will be valid in early universe cosmology.  

It should be noted that when the mediator mass $m_\phi$ becomes smaller than the inverse size of the star, $R^{-1}$, in Eq. (\ref{eq:c3}) $m_\phi^2$ in the denominator should be replaced by $R^{-2}$.  Thus, increasing the effective mass of the neutrino by going to extremely low mass of $\phi$ is not possible.  We shall be interested in $m_\phi \sim 10^{-14}$ eV, which is roughly the inverse size of red giant stars.

We recall that horizontal branch stars have core temperature of order 10 keV, radius of $5 \times 10^4$ km and density of $10^4$ g/cc.  Red giants have core temperature of order 10 keV, radius of $10^4$ km and density of $10^6$ g/cc.  Thus, $R^{-1} = 2 \times 10^{-14}$ eV for the case of red giants.  Using $m_\phi = 2 \times 10^{-14}$ eV, $y_e = 5 \times 10^{-30}$, $y_\nu = 2 \times 10^{-7}$, we obtain from the first of Eq. (\ref{eq:c3}) the effective mass of the neutrino inside red giants to be 12 MeV, which is essentially the largest value of the induced neutrino mass can have, consistent with other constraints.  Here, as shown in Ref. \cite{Babu:2019iml}, $y_e$ is bounded by electron $g-2$ ($y_e < 3.4 \times 10^{-6}$), fifth force experiments  ($y_e < 10^{-24})$, and by supernova neutrino observations ($y_e < 5 \times 10^{-30}$ -- as any larger value would make $m_\nu^{\rm eff} > 5$ MeV, with a fixed $y_\nu$).  The choice of $m_\phi = 2 \times 10^{-14}$ eV coincides with the inverse size of red giants.  The limit $y_\nu < 2 \times 10^{-7}$ arises from structure formation, which would be modified by neutrino self interactions via a light scalar \cite{Forastieri:2019cuf}. The effective number of neutrino species for big bang nucleosynthesis will increase by 0.57, which appears to be not excluded by the Planck data \cite{Aghanim:2018eyx}.

Since the induced mass of the neutrino inside red giants can be as large as 12 MeV, plasmon decays would be highly suppressed.  We could also consider interactions of $\phi$ with the nucleon instead of the electron.  In this case, the supernova limit on the coupling is $y_N < 10^{-32}$, which would lower the induced mass of the neutrinos to about 23 keV, which may still be sufficient to suppress plasmon decays into neutrinos in red giants.

With the choice of parameters that induces an in-medium mass of order 12 MeV inside red giants, the neutrino would acquire keV mass inside the Sun.  Since the solar core temperature is about a keV, and since solar neutrinos have been detected, it is necessary to require $m_\nu^{\rm eff} <$ keV in the Sun.  We note that the parameters can be chosen such that $m_\nu^{\rm eff}$ inside red giants is a 1000 times smaller, say around 12 keV, in which case $m_\nu^{\rm eff}$ inside sun would be about 2 eV.  This may affect neutrino signals from the sun, but if the new couplings are flavor universal, the medium induced mass would provide an overall phase and not affect oscillations.  The derived value of $\Delta m^2_{21}$ may be interpreted as $\Delta m^2_{21} + 2 m_0 (m_2-m_1)$, where $m_0$ is the flavor universal medium-induced mass.  If the two neutrino masses $m_1$ and $m_2$ are sufficiently close, there would be no significant departure in the determination of $\Delta m^2_{21}$ from solar neutrino and terrestrial neutrinos. 

%%%%%%%%%%%%%%%%%%%%%%%%%%%%%%%%%%%%%%%%%%%%%%%
%%%%%%%%%%%%%%%%%%%%%%%%%%%%%%%%%%%%%%%%%%%%%%%
\section{ Summary and Conclusions }\label{SEC-09}
%%%%%%%%%%%%%%%%%%%%%%%%%%%%%%%%%%%%%%%%%%%%%
%%%%%%%%%%%%%%%%%%%%%%%%%%%%%%%%%%%%%%%%%%%%%

We have revived and proposed a simplified model based on $SU(2)_H$ horizontal symmetry that can generate large neutrino transition magnetic moment without inducing unacceptably large neutrino masses.  In the $SU(2)_H$ symmetric limit, the transition magnetic moment is nonzero, while the neutrino mass vanishes.  The simplification we suggest is based on the symmetry being approximate.  

The model presented can explain the recently reported excess of electron recoil events by the XENON1T collaboration \cite{Aprile:2020tmw}.  We have explored other phenomenological consequences of the model relevant for the LHC.  We found that the prospects for detecting neutral scalar bosons decaying to $\ell^+ \tau^-$ are high in the high luminosity LHC.  The model also predicts charged scalar bosons which could lead to trilepton signatures.  

We also investigated a spin symmetry mechanism that can generate large $\mu_\nu$ while keeping $m_\nu$ small.  An example of such models is the Zee model of neutrino masses. However, we found that the value of $\mu_\nu$ induced in these models turns out be about $(2-4) \times 10^{-12}\mu_B$, which is insufficient to explain the XENON1T anomaly.

A neutrino transition magnetic moment of order $3 \times 10^{-11} \mu_B$, as needed for the XENON1T excess, would be in apparent conflict with astrophysical arguments on stellar cooling, which sets a constraint on $\mu_\nu < 4.5 \times 10^{-12} \mu_B$.  We have proposed a mechanism to evade this constraint based on interactions of neutrinos with a light scalar.  Such interactions can induce a medium dependent mass for the neutrino in the interior of stars, which could prevent kinematically energy loss by plasmon decay into neutrinos.

%%%%%%%%%%%%%%%%%%%%%%%%%%%%%%%%%%%%%%%%%%%%%
%%%%%%%%%%%%%%%%%%%%%%%%%%%%%%%%%%%%%%%%%%%%%
\section*{Acknowledgments} We thank Evgeny Akhmedov for discussions and useful comments.  
The work of KSB was supported in part by US Department of Energy Grant Number DE-SC 0016013. 

%%%%%%%%%%%%%%%%%%%%%%%%%%%%%%%%%%%%%%%%%%%%%
%%%%%%%%%%%%%%%%%%%%%%%%%%%%%%%%%%%%%%%%%%%%%
\bibliographystyle{utphys}

\bibliography{reference}

\end{document}